\documentclass[apsrev4-1,prb,superscriptaddress,longbibliography,twocolumn]{revtex4-1}
\usepackage{color}
\usepackage{amsfonts,amsmath,amssymb}
\usepackage[dvipsnames]{xcolor}
\usepackage{tabularx,hhline,tikz,multirow,array}
\usepackage{bm,enumitem}
\usepackage{dcolumn,stmaryrd}
\usepackage{comment}
\usepackage{float} % Import the float package for [H] placement specifier
\usepackage[bookmarksnumbered,pdfpagelabels=true,plainpages=false,colorlinks=true,linkcolor=red,citecolor=red,urlcolor=red]{hyperref}
\usepackage[normalem]{ulem}
\def\uv {{\bf u}}
\def\rv {{\bf r}}
\def\qv {{\bf q}}
\def\Sv {{\bf S}}
\def\Dv {{\bf D}}

\allowdisplaybreaks

\begin{document}

\title{Skyrmion and meron phases induced by spin-phonon coupling}

\author{E. Iroulart}
\affiliation{Instituto de F\'isica de L\'iquidos y Sistemas Biol\'ogicos (IFLYSIB), UNLP-CONICET, Facultad de Ciencias Exactas, Universidad Nacional de La Plata, 1900 La Plata, Argentina}
\affiliation{Departamento de F\'isica, Facultad de Ciencias Exactas, Universidad Nacional de La Plata, 1900 La Plata, Argentina}

\author{F. A. G\'omez Albarrac\'in}
\affiliation{Instituto de F\'isica de L\'iquidos y Sistemas Biol\'ogicos (IFLYSIB), UNLP-CONICET, Facultad de Ciencias Exactas, Universidad Nacional de La Plata, 1900 La Plata, Argentina}
\affiliation{Departamento de Cs. B\'asicas, Facultad de Ingenier\'ia, Universidad Nacional de La Plata, 1900 La Plata, Argentina}

\author{H. Diego Rosales}
\email{rosales@fisica.unlp.edu.ar}
\affiliation{Instituto de F\'isica de L\'iquidos y Sistemas Biol\'ogicos (IFLYSIB), UNLP-CONICET, Facultad de Ciencias Exactas, Universidad Nacional de La Plata, 1900 La Plata, Argentina}
\affiliation{Departamento de Cs. B\'asicas, Facultad de Ingenier\'ia, Universidad Nacional de La Plata,  1900 La Plata, Argentina}

\date{\today}

\begin{abstract}
In chiral magnets, magnetic skyrmions are typically stabilized by the competition between exchange and Dzyaloshinskii-Moriya (DM) interactions under an external magnetic field, while the role of lattice degrees of freedom has received comparatively less attention. Here we study how spin-phonon (SP) coupling modifies magnetic interactions and the resulting spin textures in a two-dimensional skyrmion model in the square lattice. Using Monte Carlo simulations, we compare two simplified models describing the SP coupling: the Einstein site-phonon (ESP) and bond-phonon (BP) models.  Integrating the lattice degrees of freedom, we obtain effective Hamiltonians for each SP model, which involve three-spin interactions and quadratic terms. In both cases, skyrmion crystals are stabilized in field regimes that are topologically trivial in the uncoupled model. Moreover, in certain regimes, the conventional triple-$\qv$ hexagonal skyrmion lattice is distorted into a double-$\qv$ square skyrmion lattice. Additionally, we find that, in the ESP model, strong phonon coupling affecting the exchange interactions drives the system from typical low-field helical phases to meron-antimeron crystals with no net chirality. Overall, our results show that lattice effects provide a simple mechanism to tune a wide variety of topological magnetic phases, ranging from typical skyrmion crystals to distorted arrangements of skyrmions, elongated bimerons and meron-antimeron phases.
\end{abstract}

\maketitle
%%%%%%%%%%%%%%%%%%%%%
\section{Introduction}

Magnetic skyrmions are topological spin textures that emerge from the competition between different magnetic interactions and have been widely studied in a variety of systems \cite{yu2012skyrmion,al2001skyrmions,yi2009skyrmions,rossler2010skyrmionic,gobel2020beyond,zhang2023magnetic}. Their formation is typically associated with the interplay between exchange and Dzyaloshinskii-Moriya (DM) interactions \cite{dzyaloshinsky1958thermodynamic,moriya1960anisotropic}, although other mechanisms such as exchange frustration \cite{rosales2015three,hayami2016bubble}, bond-dependent anisotropies \cite{yi2009skyrmions,gao2020fractional,amoroso2020spontaneous,rosales2022anisotropy,dos2026tunable}, long-range interactions \cite{utesov2021dipolar,rkkyskyrmion2018}, and higher-order exchange terms \cite{okubo2012multiple,mohylna2022spontaneous} can also stabilize skyrmion phases.

At low temperatures and zero magnetic field, these systems often favor helical states characterized by one-dimensional spin modulations. As the magnetic field increases, the competition between interactions leads to multi-$\qv$ states and the formation of skyrmion crystals (SkX). In intermediate regimes, additional textures such as elongated skyrmions or bimerons can appear, reflecting the presence of competing energy scales. Beyond skyrmions, related topological textures such as merons have been observed in different physical contexts \cite{yu2018transformation,moon1995spontaneous,guerci2022designer,augustin2021properties,mohylna2025frustration,casas2023coexistence,lin2015skyrmion,guo2020meron,wang2021meron}. These objects carry fractional topological charge and typically appear in pairs, with an energetics similar to that of XY vortices.

While most studies focus on magnetic interactions alone, real materials also involve lattice degrees of freedom. Disorder and lattice distortions are known to modify the stability of skyrmion phases \cite{diaz2017fluctuations,mirebeau2018spin,hoshino2018theory,chudnovsky2018skyrmion,reichhardt2022statics,liu2022disorder,henderson2022skyrmion,silva2014emergence,mohylna2023robustness,rosales2024robustness,bo2024suppression}. In this context, spin-phonon coupling provides a natural way to tune magnetic interactions. Such coupling has been extensively studied in frustrated and correlated systems, including spin-ice materials \cite{gomez2013spin,borzi2016intermediate}, kagome lattices\cite{wang2008spin,gen2022nematicity}, and pyrochlore systems \cite{aoyama2021effects,bergman2006models}. However, its role in stabilizing skyrmion textures and related topological phases remains less explored. Recently, it has been shown that the magnon-phonon interaction from out-of-plane phonon modes may induce skyrmion and antiskyrmion textures in a simple ferromagnetic model under a magnetic field \cite{go2026phonon}.

Here, we investigate how spin-phonon coupling alters the effective magnetic interactions and stabilizes new textures in a two-dimensional skyrmion model on a square lattice. We consider two coupling schemes: the Einstein site-phonon model \cite{bergman2007degenerate,wang2008spin,pili2019two,gomez2013spin}, where lattice distortions generate effective multi-spin interactions, and the bond-phonon model \cite{penc2004half,nascimento20242d,bergman2006models}, where the spin-phonon coupling acts locally on each bond. By analyzing how these distortions affect the exchange and DM interactions,  we obtain two effective Hamiltonians with three spin interactions and quadratic terms. Through extensive Monte Carlo simulations, we explore the emergence of skyrmion, meron, and mixed topological phases, focusing on the low-temperature regime. Our main goal is to show how the spin-phonon coupling applied to a typical skyrmion model gives rise to a plethora of non-trivial topological phases.

The rest of the paper is organized as follows. In Sec.~\ref{sec:Model}, we introduce the models and numerical methods. In Sec.~\ref{sec:results}, we present the results. Section~\ref{sec:conclusions} summarizes our findings.

%%%%%%%%%%%%%%%%%%%%%%%%%%%%%
\section{Model and Method}
\label{sec:Model}
%%%%%%%%%%%%%%%%%%%%%%%%%%%%%

%
\begin{figure}[h!] % Use [H] for precise placement
\centering
\includegraphics[width=0.8\columnwidth]{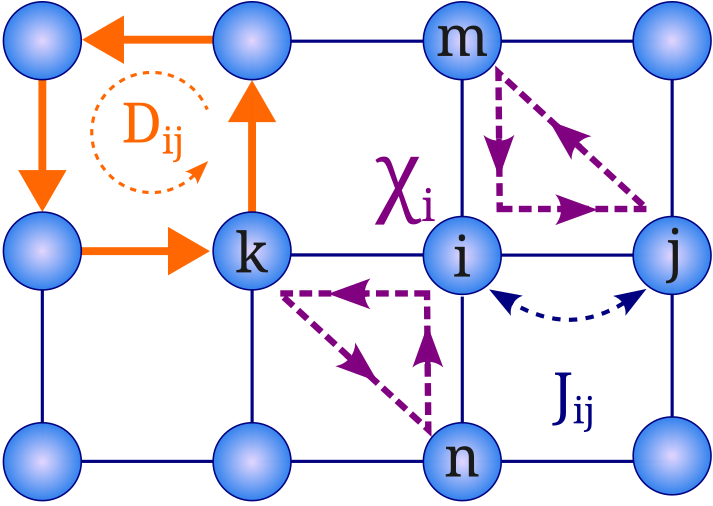}
\caption{Illustration of the square lattice, indicating nearest neighbors exchange  $J_{ij}$ between sites $(i,j)$ (blue) and Dzyaloshinskii-Moriya interaction $\Dv_{ij}$. Purple dashed arrows indicate the orientation used to compute the scalar chirality (Eq.~\ref{Chirality}) from spins at sites $(i,j,m)$.
}
\label{chirality_lattice}
\end{figure}
We consider the classical ferromagnetic Heisenberg model on a square lattice:
\begin{equation}
\small
H_{mag} = \sum_{\langle ij \rangle} J_{ij} \, \Sv_i \cdot \Sv_j 
+ \Dv_{ij} \cdot (\Sv_i \times \Sv_j) - B\sum_i S^z_i,
\label{Hamiltonian1}
\end{equation}

\noindent where the spin variables $\Sv_i$ are classical unit vectors at site $\textbf{r}_i$, $J_{ij}$ is the exchange interaction, $\Dv_{ij}= D_{ij}\,(\textbf{r}_j-\textbf{r}_i)/\lvert\textbf{r}_j-\textbf{r}_i\rvert ={D}_{ij}\,\hat{\textbf{r}}_{ij}$ is the DM interaction, with amplitude $D_{ij}$ between neighboring sites $(i,j)$, $B$ is the strength of the magnetic field along the $z$ axis, and the sum $\sum_{\langle i,j \rangle}$ is taken over nearest neighbors.

To include magnetoelastic coupling, we allow the exchange and DM interactions to depend linearly on atomic displacements. Let $\uv_i$ denote the displacement of site $i$ from its equilibrium position $\rv^0_{i}$. For small distortions $|\uv_i|/|\rv^0_{i}|<<1$, the interactions can be expanded as \cite{aoyama2021effects}:
\begin{equation}
J_{ij} = -J_0 \left(1 + \alpha\, u_{ij} \right), \quad
\Dv_{ij} = \Dv_{0ij} \left(1 + \beta\, u_{ij} \right),
\label{ExpandedInteractions}
\end{equation}

where we take $J_0>0$ for ferromagnetic interactions and $\Dv_{0ij}= {D}_{0}\,\hat{r}_{ij}$. Here, $u_{ij}$ is the change in distance between neighboring spins
$\mathbf S_i$ and $\mathbf S_j$ relative to the equilibrium bond length.
Throughout this work, we assume that lattice distortions modify only the magnitude of the DM interaction, while its direction remains unchanged. Moreover, we restrict the displacements to the lattice plane, as opposed to previous works \cite{go2026phonon}.

The parameters $\alpha$ and $\beta$ entering Eq.~(2) quantify the
relative sensitivity of the exchange and Dzyaloshinskii--Moriya
interactions to changes in the bond length. They can be expressed as
the logarithmic derivatives
\begin{equation}
\alpha=
\left.
\frac{1}{J_0}\frac{dJ(r)}{dr}
\right|_{r=r_0},
\qquad
\beta=
\left.
\frac{1}{D_0}\frac{dD(r)}{dr}
\right|_{r=r_0},
\label{Eq:alphabeta}
\end{equation}
where $r_0=|\mathbf r_{ij}^{0}|$ is the equilibrium nearest-neighbor
bond length. After integrating out the
phonon degrees of freedom, only the rescaled combinations
$\alpha/\sqrt{\kappa}$ and $\beta/\sqrt{\kappa}$ enter the effective
Hamiltonians of Eqs.~(5) and (7). Consequently, their numerical value
cannot be compared one-to-one with tabulated microscopic
exchange-striction or DM-striction coefficients without also
specifying the elastic stiffness $\kappa$, which is material
dependent and not fixed by our model.

A $\kappa$-independent, physically transparent measure of the coupling
strength is instead the ratio between the phonon-induced quadratic
term and the corresponding bare bilinear interaction. For the DM
sector, for instance,
\begin{equation}
r_\beta \equiv \frac{\beta^2 D_0^2}{2\kappa J_0}
        = (\beta')^2\,\frac{D_0^2}{2 J_0},
\end{equation}
where $\beta' = \beta/\sqrt{\kappa}$ is the rescaled coupling used
throughout Sec.~III. For $D_0/J_0 = 1$, as employed in our
simulations, the explored range $\beta' \lesssim 0.8$ corresponds to
$r_\beta \lesssim 0.3$, i.e., phonon-induced corrections of at most
$\sim 30\%$ of the leading exchange energy scale. An analogous
estimate holds for $\alpha'$ and the exchange sector.

Magnetoelastic couplings of comparable, and in some cases larger,
relative magnitude have been reported in frustrated magnetic
insulators with sizable spin-lattice coupling, such as spin-ice and
kagome and pyrochlore compounds~\cite{bergman2006models,wang2008spin,gomez2013spin,borzi2016intermediate}.
An exchange-mediated magnetoelastic coupling of comparable relative strength has also been estimated for yttrium iron garnet~\cite{streib2019magnon}. Although that work considers dynamical magnon-phonon interactions rather than the quasi-static phonon models studied here, it provides a useful reference for the magnitude of exchange-mediated magnetoelastic couplings.

On this basis, we regard the strong-coupling portion of our phase
diagrams ($\alpha',\beta'\gtrsim 0.5$) as representative of materials
with pronounced magnetoelastic response, rather than as a generic
weak-coupling regime. The topological textures stabilized in this
regime, including meron-antimeron, compact meron-antimeron, and square
double-$\mathbf q$ skyrmion crystals (see Sec.~III A for definitions),
should accordingly be viewed as predictions relevant to strongly
magnetoelastic chiral magnets, to be tested against material-specific
estimates of the exchange- and DMI-striction couplings.

%%%%%%%%%%%%%%%%%%%
\subsection{Einstein site-phonon model }
%%%%%%%%%%%%%%%%%%%

In the ESP model, each site possesses an independent displacement vector $\uv_i$ from its equilibrium position $\rv_i^0$ and $u_{ij}$ is given by $u_{ij} = \hat{e}_{ij} \cdot (\uv_i - \uv_j)$, where $\hat{e}_{ij} = (\rv_j^0 - \rv_i^0)/|\rv_j^0 - \rv_i^0|$ is the unit vector along the bond direction. The phonons are treated as local harmonic oscillators \cite{jia2005lattice}, with an elastic energy given by
\begin{equation}
    H_{\mathrm{ph}}^{\text{ESP}} = \frac{1}{2} \kappa \sum_i |\uv_i|^2, 
\end{equation}
where $\kappa$ is the elastic constant. 

Since the elastic term is quadratic and the coupling to spins is linear, the phonon degrees of freedom can be integrated out exactly. In the classical limit, this is equivalent to minimizing the total Hamiltonian $H_{\text{tot}}^{\text{ESP}}=H_{\text{mag}}+H_{\text{ph}}^{\text{ESP}}$ with respect to $\uv_i$ (i.e., $\partial H_{tot}/\partial \uv_i = 0$), yielding
\begin{equation}
\kappa\, \uv^*_i = \sum_{j \in \text{NN}(i)} \hat{e}_{ij} \left[
J_0 \alpha\, \Sv_i \cdot \Sv_j - \beta\, \Dv_{0ij} \cdot (\Sv_i \times \Sv_j)
\right].
\end{equation}
This expression provides the optimal displacement $\uv_i^*$ at each site as a function of the local spin configuration. Replacing $\uv_i^*$ back into the Hamiltonian yields an effective spin Hamiltonian $H_{\mathrm{eff}}^{\text{ESP}}$ that includes induced multi-spin interactions arising from the coupling to the lattice. This Hamiltonian reads

\begin{widetext}
\begin{eqnarray}
H_{\mathrm{eff}}^{\text{ESP}} &=& \sum_{\langle ij \rangle} \biggl[-J_0 \Sv_i \cdot \Sv_j + \Dv_{0ij} \cdot (\Sv_i \times \Sv_j) \nonumber\\
&-& \frac{1}{2\,\kappa} \biggl( J_0^2 \alpha^2 (\Sv_i \cdot \Sv_j)^2 +  \beta^2 \left( \Dv_{0ij} \cdot (\Sv_i \times \Sv_j) \right)^2 - 2\,J_0\,\alpha\,\beta (\Sv_i \cdot \Sv_j) \left( \Dv_{0ij} \cdot (\Sv_i \times \Sv_j) \right)\biggr) \biggr] \nonumber\\
&+& \sum_i \sum_{\substack{j \neq k \\0 j,k \in \mathrm{NN}(i)}}\biggl[ -\frac{1}{2\,\kappa} (\hat{e}_{ij} \cdot \hat{e}_{ik}) \biggl( J_0^2 \alpha^2 (\Sv_i \cdot \Sv_j)(\Sv_i \cdot \Sv_k)+ \beta^2 \left( \Dv_{0ij} \cdot (\Sv_i \times \Sv_j) \right) \left( \Dv_{0ij} \cdot (\Sv_i \times \Sv_k) \right)\nonumber\\
&-&2J_0\,\alpha\,\beta \left[ (\Sv_i \cdot \Sv_j)\left( \Dv_{0ij} \cdot (\Sv_i \times \Sv_k) \right) \right] \biggr) \biggr] - B\sum_i S^z_i.
\label{HamiltonianEff1}
\end{eqnarray}
\end{widetext}

The multi-spin and further-neighbor interactions terms in Eq.~(\ref{HamiltonianEff1}) depend on the relative orientation of the bonds $\hat{e}_{ij}$ and introduce competing tendencies controlled by $\alpha$ and $\beta$. Among these contributions, the quadratic terms $(\mathbf{S}_i\cdot\mathbf{S}_j)^2$ and $(\mathbf{D}_{0ij}\cdot(\mathbf{S}_i\times \mathbf{S}_j))^2$ do not distinguish between positive and negative spin correlations, nor between opposite chiralities, leading to local degeneracies. In contrast, the three-site terms couple neighboring bonds through geometric factors. The contributions proportional to $\alpha^2$ favor configurations where adjacent spin correlations have opposite sign, promoting locally antiferromagnetic patterns along lattice directions. Similarly, the $\beta^2$ terms favor alternating chirality between neighboring bonds, disfavoring uniform rotational textures.

%%%%%%%%%%%%%%%%%%%
\subsection{Bond phonon model}
%%%%%%%%%%%%%%%%%%%

In contrast, in the BP model~\cite{penc2004half,motome2006monte,bergman2006models}, the lattice degrees of freedom are assigned to the bonds rather than to the sites. Each nearest-neighbor link $\langle ij \rangle$ carries an independent longitudinal displacement $u_{ij}$, and the corresponding elastic energy is
\begin{equation}
    H_{\mathrm{ph}}^{\mathrm{BP}} = \frac{1}{2} \kappa \sum_{\langle ij \rangle} u_{ij}^2,
\end{equation}

\noindent where $\kappa$ is the bond elastic constant. As in the ESP model, the exchange and DM interactions are taken to depend linearly on $u_{ij}$. Integrating out the phonon variables in the combined Hamiltonian $H_{\text{tot}}^{\text{BP}}=H_{\text{mag}}+H_{\text{ph}}^{\text{BP}}$ yields the effective spin model
\begin{equation}
\begin{split}
    H_{\mathrm{eff}}^{\mathrm{BP}} = \sum_{\langle ij \rangle} & \biggl[-J_0 \Sv_i \cdot \Sv_j +\Dv_{0ij} \cdot (\Sv_i \times \Sv_j) \\
    &- \frac{J_0^2 \alpha^2}{2 \kappa} (\Sv_i \cdot \Sv_j)^2 
    - \frac{\beta^2}{2 \kappa} \left( \Dv_{0ij} \cdot (\Sv_i \times \Sv_j) \right)^2 \\
    &+ \frac{J_0 \alpha \beta}{\kappa} (\Sv_i \cdot \Sv_j) \left[ \Dv_{0ij} \cdot (\Sv_i \times \Sv_j) \right] \biggr]- B\sum_i S^z_i.
\end{split}
\label{HamiltonianEff2}
\end{equation}
\noindent The BP Hamiltonian contains only on-bond corrections: all additional terms involve the two spins connected by a given link. No geometric factors couple different bonds, and therefore no further-neighbor or multi-spin interactions are generated. It is important to mention that, since displacements are considered as independent variables, this model cannot be interpreted as a fully realistic description of lattice deformations generated by external strain or pressure. Instead, it should be viewed as a simplified phenomenological model that captures local magnetoelastic effects while neglecting compatibility constraints and long-range elastic correlations.

In this case, the roles of $\alpha$ and $\beta$ can be understood directly from the local quadratic terms. The contribution proportional to $\alpha^2$ enhances $(\mathbf{S}_i\cdot\mathbf{S}_j)^2$, favoring collinear alignment without distinguishing between parallel and antiparallel configurations. This acts uniformly on all bonds and tends to reinforce locally aligned or polarized states. On the other hand, the $\beta^2$ term increases the weight of $(\mathbf{D}_{0ij}\cdot(\mathbf{S}_i\times\mathbf{S}_j))^2$,  which combined with the typical DM term, favors chiral textures.

As a result, spiral or skyrmion-like textures can be stabilized when allowed by the underlying interactions.

\subsection{Monte Carlo Simulations}
\label{sec:MCsimulations}

The simulations were carried out using the Metropolis Monte Carlo (MC) algorithm.  We considered square lattices of $N=L^2$ spins, and linear size $L=32, 48,$ and $64$, under periodic boundary conditions. The system was cooled using an exponential schedule from high temperature through $80$ successive steps down to $T/J_0=1.2\times 10^{-3}$, taking $10^5$ MC steps for initialization and four times as many for measurements. The annealing technique is chosen to explore thermodynamically stable phases. For each set of parameters, up to $20$ copies were run starting from random configurations at high temperature. This procedure allows us to check the robustness of the low-temperature phases against different independent initial conditions.

To characterize the various types of chiral phases, we computed the local scalar chirality $\chi_i$, defined as
\begin{equation}
    \chi_i = \Sv_{i} \cdot \left( \Sv_{j} \times \Sv_{m} + \Sv_{k} \times \Sv_{n} \right),
    \label{Chirality_local}
\end{equation}
where $\{\Sv_{i},\Sv_{j},\Sv_{m}\}$ ($\{\Sv_{i},\Sv_{k},\Sv_{n}\}$) are three spins in an up (down) triangular plaquette, as depicted in Fig.~\ref{chirality_lattice}.  Geometrically, \(\chi_i\) is related to the solid angle formed by the three spins in the plaquette~\cite{aoyama2022emergent}, vanishing for coplanar configurations and becoming finite for chiral spin textures. The total scalar chirality of the system is then obtained by averaging the local scalar chirality over all plaquettes:
\begin{equation}
    \chi = \frac{1}{4\pi}  \left\langle \chi_i \right\rangle= \frac{1}{4\pi}  \left\langle \sum_{i}\Sv_{i} \cdot \left( \Sv_{j} \times \Sv_{m} + \Sv_{k} \times \Sv_{n} \right) \right\rangle,
    \label{Chirality}
\end{equation}
Chirality $\chi$ is the discrete version of the topological charge $Q = \frac{1}{4\pi} \iint \Sv \cdot \left( \frac{\partial \Sv}{\partial x} \times \frac{\partial \Sv}{\partial y} \right) dx\,dy $, which accounts for the magnetization configuration wrapping around the unit sphere. 
For skyrmions, $Q=-1$ \cite{yin2016topological}, while helical and ferromagnetic phases yield $Q=0$.

Additionally, we computed the in-plane structure factor $S_{\perp}(\qv)$
\begin{equation}
S_{\perp}(\qv) = \frac{1}{N} \left\langle \left| \sum_j S^x_{j} e^{i\,\qv\cdot \textbf{r}_j} \right|^2 + \left| \sum_j S^y_{j} e^{i\,\qv\cdot \textbf{r}_j} \right|^2 \right\rangle
\label{SFxy}
\end{equation}
which allows us to distinguish between helical, skyrmion-crystal, and fully polarized phases. A hexagonal skyrmion crystal corresponds to a triple-$\qv$ state formed by the superposition of three helical modulations satisfying $\sum_{i=1}^3 \qv_i = \textbf{0}$ \cite{matsumura2024single}. This leads to six characteristic Bragg peaks in the structure factor, in agreement with neutron scattering experiments \cite{muhlbauer2009skyrmion,jonietz2010spin,munzer2010skyrmion,yu2010real}.

In contrast, crystals formed by merons and antimerons exhibit a square arrangement that produces four dominant peaks in reciprocal space \cite{yu2018transformation}. This distinct signature reflects the underlying two-by-two periodicity of these topological defects. The different peak structures in reciprocal space therefore reflect the symmetry and ordering wave vectors of the corresponding magnetic textures.
%The structure factor therefore provides a useful fingerprint for identifying the ordering wave vectors and distinguishing between different chiral textures.

In the following section, we analyze the effects of the SP couplings, characterized by $\alpha$ and $\beta$.
\begin{figure*}[th!]
\includegraphics[width=0.8\textwidth]{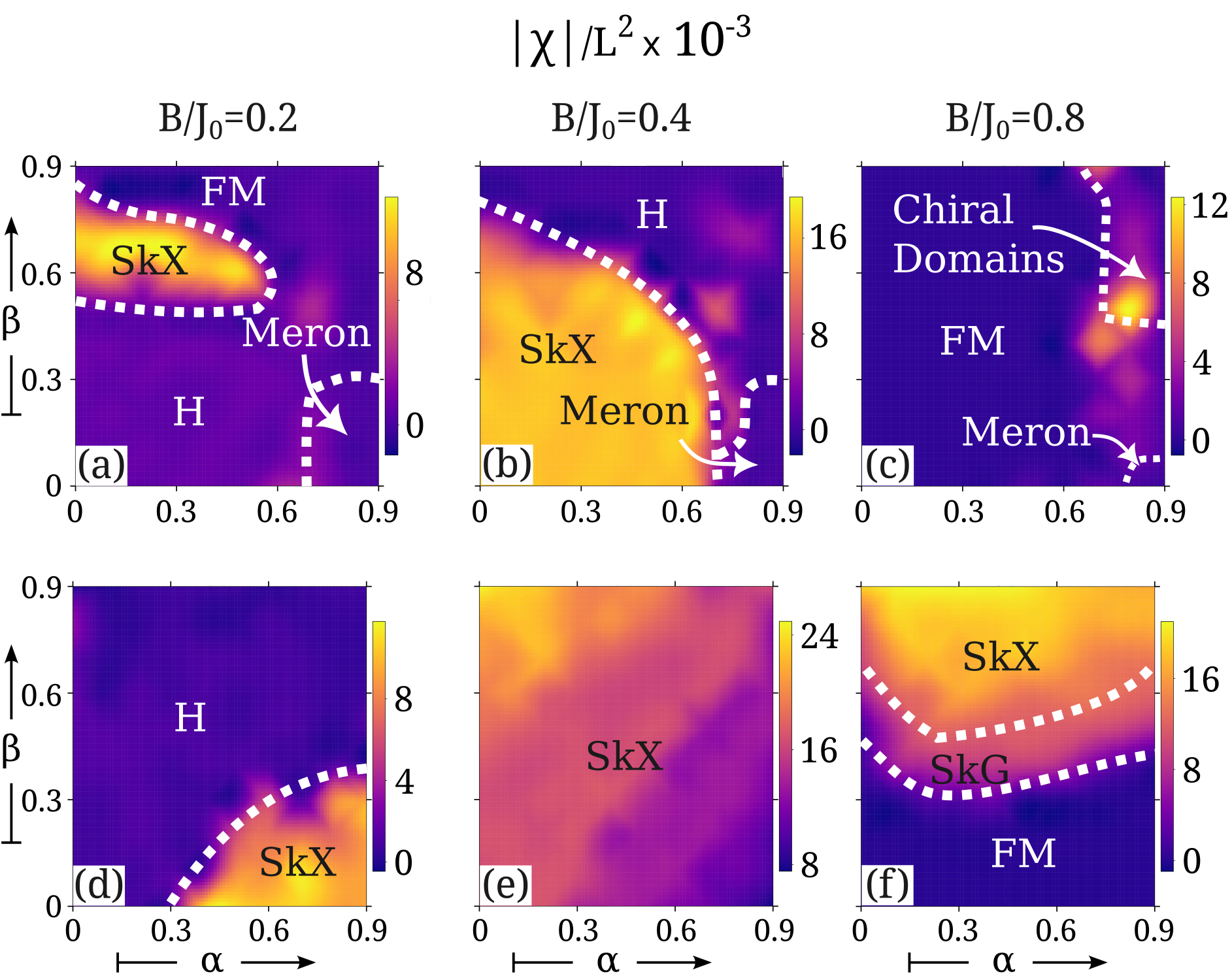}
\caption{(Color online) Density plots of the absolute normalized chirality $|\chi|/L^2$ as a function of $\alpha$ and $\beta$ for magnetic fields $B/J_0=0.2, 0.4$ and $0.8$ at the lowest simulated temperature $T/J_0 = 1.2 \times 10^{-3}$. Top row: Einstein site-phonon model. Bottom row: Bond-phonon model. The different phases are indicated: helices (H), skyrmion crystals (SkX), skyrmion gas (SkG), meron textures (Meron) and ferromagnetic order (FM).}
\label{fig:PDChir}
\end{figure*}
%

%%%%%%%%%%%%%%%%%%%%%
\section{Results}% for $\alpha and beta>0$}
\label{sec:results}
%%%%%%%%%%%%%%%%%%%%%

We now study the effects of SP coupling on the magnetic phases of the skyrmion model through simulations of the effective Hamiltonians introduced above. To simplify the analysis, we rescale the phonon displacements and coupling parameters as $\mathbf{u} \rightarrow \frac{\mathbf{u}}{\sqrt{\kappa}}$, $\alpha \rightarrow \alpha' = \alpha / \sqrt{\kappa}$, and $\beta \rightarrow \beta' = \beta / \sqrt{\kappa}$. 
This absorbs the elastic constant into the couplings and reduces the number of independent parameters. In the following, we denote the rescaled parameters simply as $\alpha$ and $\beta$.

\begin{figure*}[ht!]
\includegraphics[width=0.85\textwidth]%{Snapshots_sites_bonds_B_0_0.2_0.4_0.8_axes.pdf}
{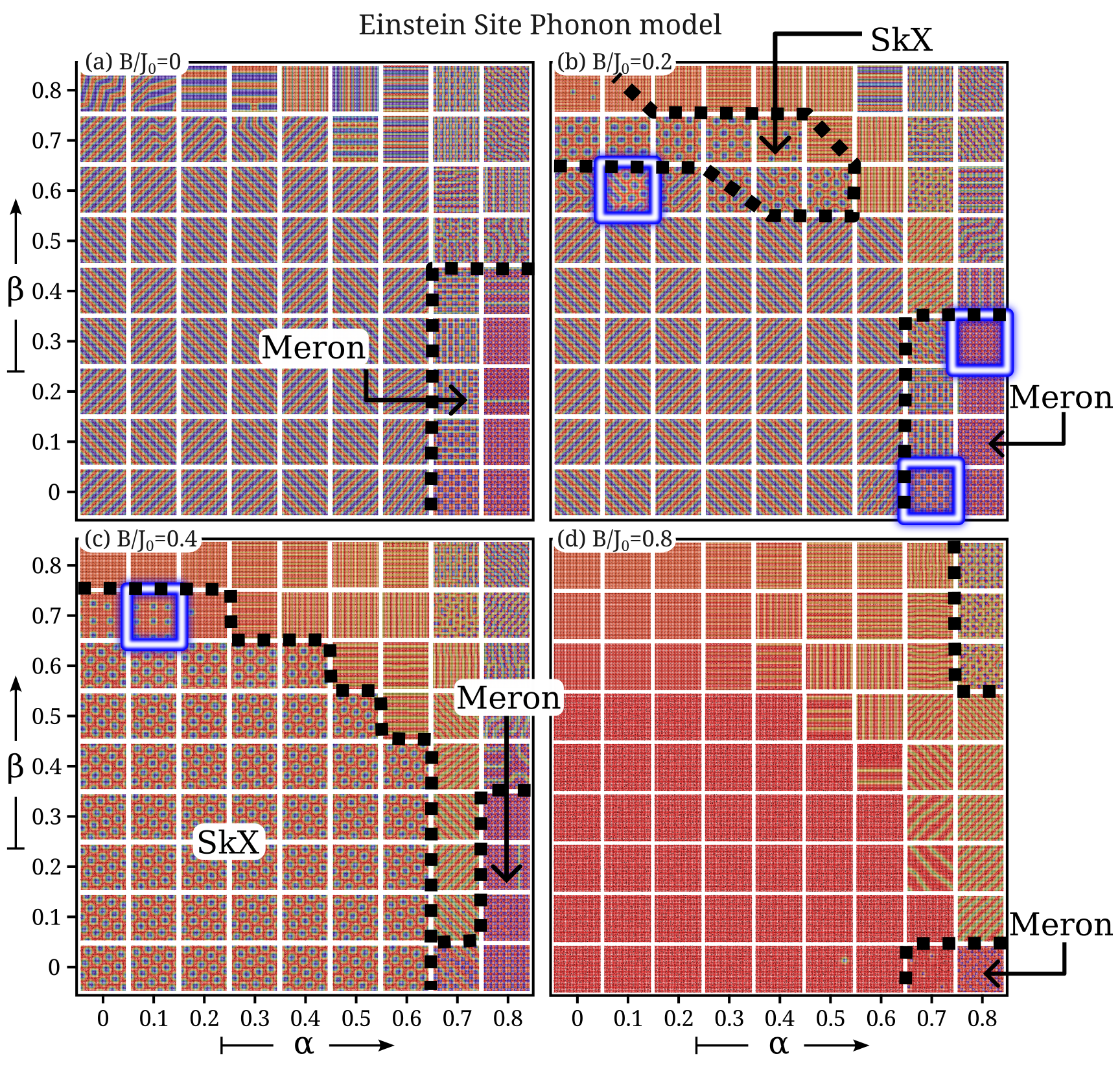}
\caption{(Color online) Representative magnetic configurations for the Einstein site-phonon model as a function of $\alpha$ and $\beta \in [0,0.8]$ at the lowest simulated temperature $T/J_0 = 1.2 \times 10^{-3}$. Panels correspond to magnetic fields (a) $B/J_0 = 0$, (b) $0.2$, (c) $0.4$, and (d) $0.8$. The grid illustrates the evolution of magnetic textures across parameter space, including helical (H), skyrmion crystal (SkX), meron–antimeron, and mixed phases. Dotted lines indicate approximate phase boundaries, to be compared with the corresponding scalar-chirality phase diagrams in Fig.~\ref{fig:PDChir} (top row). Representative snapshots highlighted by blue rectangles are shown in greater detail in Fig.~\ref{fig:SnapPhases} (middle row).
}
\label{fig:GridSnapsESP}
\end{figure*}

\begin{figure*}[ht!]
\includegraphics[width=0.85\textwidth]
{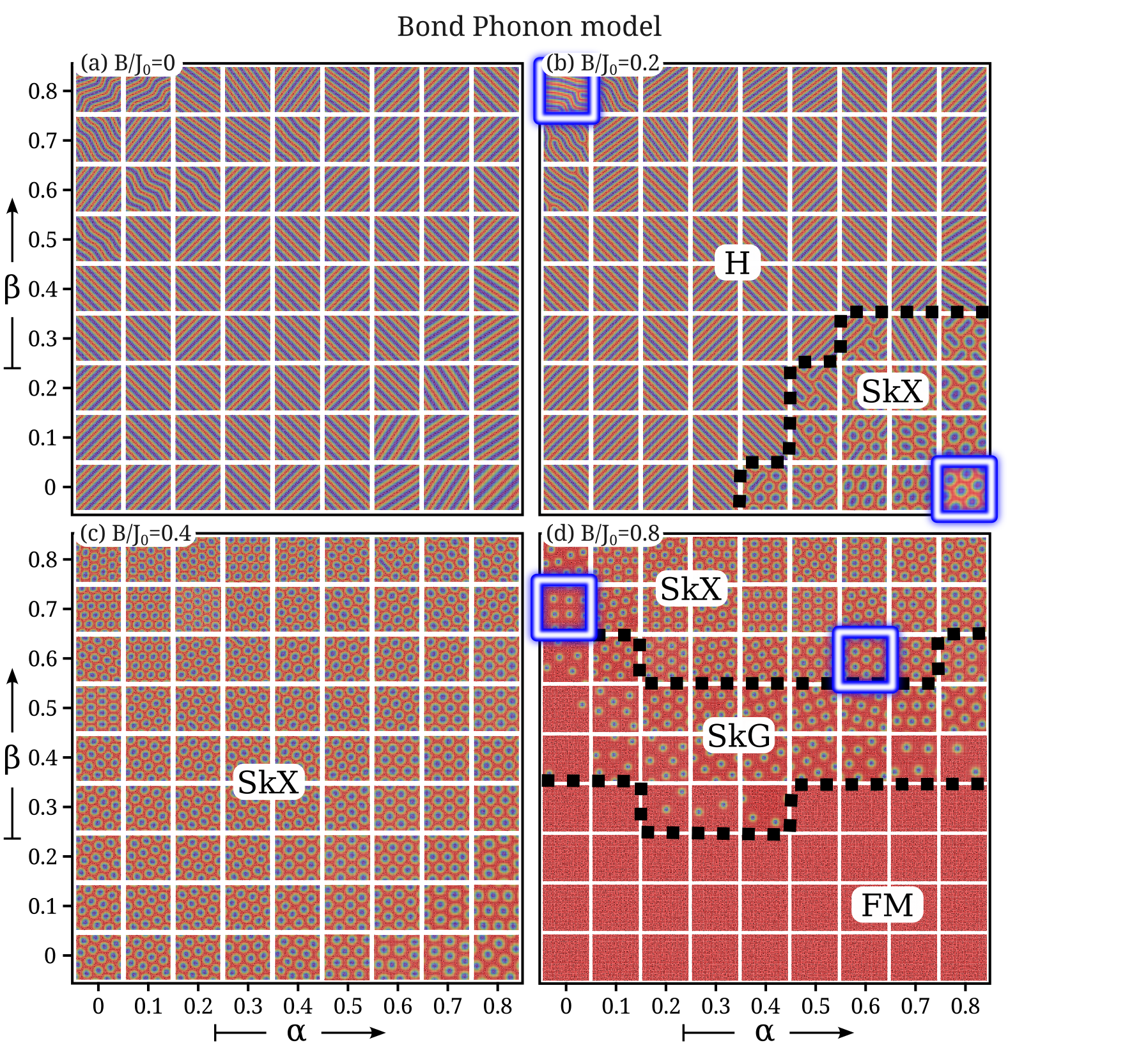}
\caption{(Color online) Representative magnetic configurations for the bond-phonon model as a function of $\alpha$ and $\beta \in [0,0.8]$ at the lowest simulated temperature $T/J_0 = 1.2 \times 10^{-3}$. Panels correspond to magnetic fields (a) $B/J_0 = 0$, (b) $0.2$, (c) $0.4$, and (d) $0.8$. Compared to the ESP model, the BP coupling predominantly stabilizes skyrmion textures over a wider region of parameter space, with fewer competing phases. Dotted lines indicate approximate phase boundaries, to be compared with the corresponding scalar-chirality phase diagrams in Fig.~\ref{fig:PDChir} (bottom row). Representative snapshots highlighted by blue rectangles are shown in greater detail in Fig.~\ref{fig:SnapPhases} (bottom row).}
\label{fig:GridSnapsBP}
\end{figure*}

We explore the resulting magnetic phases as a function of $\alpha$, $\beta$, and the external magnetic field $\mathbf{B}=(0,0,B/J_0)$. In Fig.~\ref{fig:PDChir}, the low-temperature  $\beta$ vs $\alpha$ phase diagrams for three different values of the magnetic field $B/J_0=0.2, 0.4$, and $0.8$ are shown for both ESP and BP models. These phase diagrams are constructed using the absolute normalized chirality $|\chi|/L^2$, which is a key order parameter to detect possible skyrmion phases. At first glance, we see that in both SP models, different chiral phases are stabilized as $\alpha$ and $\beta$ increase. The plots highlight the different evolution of chiral phases in the two coupling schemes, with the ESP model showing a more restricted and fragmented chiral region, while the BP model exhibits a broader region with finite chirality.
On the one hand, in the BP model, skyrmion phases are enhanced in regions where, in the model without SP coupling, non-chiral phases, such as spirals at low fields and polarized phases at higher fields, are found. On the other hand, in the ESP model, other exotic textures arise, such as merons, which we will describe below. Representative snapshots of the different phases for these phase diagrams are presented for the ESP and BP models in Figs.~\ref{fig:GridSnapsESP} and Fig.~\ref{fig:GridSnapsBP}, respectively.

\begin{figure*}[th!]
\includegraphics[width=\textwidth]{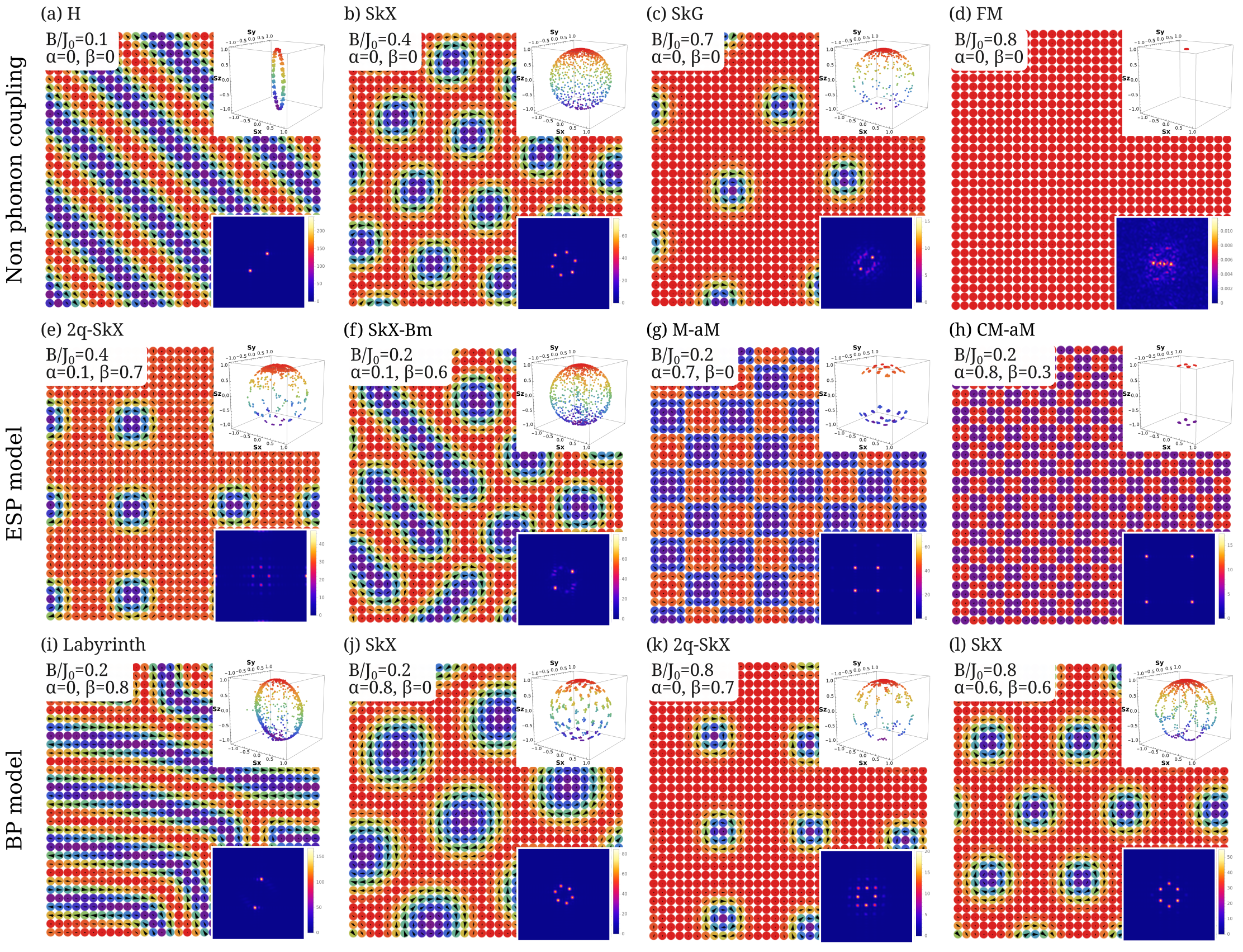}
    \caption{(Color online) Representative magnetic configurations under different magnetic fields and spin-phonon coupling schemes at the lowest simulated temperature $T/J_0 = 1.2 \times 10^{-3}$. The first row (panels (a)-(d)) shows the reference model without SP coupling ($\alpha=\beta=0$) for $B/J_0=0.1$, $0.4$, $0.6$, and $0.8$, corresponding to: (a) helical phase (H), (b) hexagonal skyrmion crystal (SkX), (c) skyrmion gas (SkG), and (d) fully polarized phase (FM). The second row (panels (e)-(h)) displays configurations in the ESP model, illustrating phases induced by SP coupling, including: (e) square double-$\qv$ skyrmion crystal (2$\qv$-SkX), (f) skyrmion–bimeron (SkX-Bm) phase, (g) meron-antimeron crystal (M-aM), and (h) compact meron-antimeron (CM-aM) phase. The third row (panels (i)-(l)) shows the corresponding configurations for the BP model. The snapshots chosen for the ESP and BP models are highlighted in Figs.~\ref{fig:GridSnapsESP} and \ref{fig:GridSnapsBP}, respectively.}
\label{fig:SnapPhases}
\end{figure*}

In the absence of a magnetic field, the BP model displays a smooth deformation of the helical phase into curved spirals, and eventually into horizontal stripes in the large-$\beta$ limit (see Fig.~\ref{fig:GridSnapsBP}(b)). In contrast, the ESP model leads to a broader variety of textures (Fig.~\ref{fig:GridSnapsESP}(a)), ranging from helices to meron-antimeron (M-aM) crystals and compact meron-antimeron (CM-aM) phases (which will be described in detail in Sec.~\ref{Subsec:Phases}) at large $\alpha$. For intermediate and large values of both $\alpha$ and $\beta$, competing patterns appear, combining spirals and meron-like structures, giving rise to labyrinth-like configurations. In the large-$\beta$ limit, vertical and horizontal spiral patterns are also observed.

For intermediate magnetic fields ($0.2\leq B/J_0 \leq 0.4$), the BP model (Fig.~\ref{fig:GridSnapsBP}(b-c)) favors skyrmion textures, expanding the region of $(\alpha,\beta)$ parameter space where chiral phases are stabilized. In contrast, in the ESP model (Fig.~\ref{fig:GridSnapsESP}(c)), skyrmion configurations are suppressed at large $\alpha$ and $\beta$. Instead, the competition between exchange and DM interaction modulations leads to phases such as a double-$\qv$ square skyrmion lattice (around $\beta\approx0.7$, $\alpha=0$), regions with coexistence of chiral domains, and meron phases at large $\alpha$. The square skyrmion lattice is also found for large $\beta$ in the BP model at larger fields, see Fig. \ref{fig:GridSnapsBP}(d).

At high magnetic field ($B/J_0=0.8$), the two models behave differently. The ESP model (Fig.~\ref{fig:GridSnapsESP}(d)) is mostly polarized, with only isolated skyrmions or residual helical structures at larger $\alpha$, leading to meron phases as the coupling is further increased. In contrast, the BP model shows the stabilization of skyrmions for $\beta\gtrsim0.8$, driven mostly by        $\beta$.  Increasing this coupling enhances the effective strength of chiral interactions, promoting non-collinear textures and stabilizing skyrmion phases over a broad range of magnetic fields. In contrast, increasing $\alpha$ enhances the sensitivity of the exchange interaction to lattice distortions, introducing competing tendencies that can destabilize skyrmion order and favor phases such as meron–antimeron crystals. In the next subsection, we describe some of these phases in more detail.

%%%%%%%%%%%%%
\subsection{Description of Magnetic Phases}
\label{Subsec:Phases}
%%%%%%%%%%%%%
Below, we characterize the different magnetic phases and relate them to the underlying competition between the effective terms of the ESP and BP models presented in Eq.(\ref{HamiltonianEff1}) and (\ref{HamiltonianEff2}), respectively.
\subsubsection{Case without spin-phonon coupling ($\alpha = \beta = 0$)}  
We first consider the non-phonon coupling case by setting $\alpha = \beta = 0$ in Eq.~(\ref{Hamiltonian1}), recovering the well-studied skyrmion model with ferromagnetic exchange and in-plane DM interactions under a magnetic field ~\cite{bogdanov1989thermodynamically,roessler2006spontaneous,yu2010real,ezawa2011compact,iroulart2024probing}.  
For $D_0/J_0 = 1$ at low temperature, the phase diagram contains three main regions: a single-$\qv$ helical (H) state for $0 \leq B/J_0 < B_{c1} \simeq 0.3$, a hexagonal skyrmion crystal (SkX) for $B_{c1} \leq B/J_0 \leq B_{c2} \simeq 0.6$, and a fully polarized ferromagnetic (FM) phase for $B/J_0 > B_{c2}$.  
Thermal fluctuations further generate an intermediate skyrmion gas (SkG) between SkX and FM phases.  
Representative configurations are shown in Fig.~\ref{fig:SnapPhases}(a)-(d), together with their structure factors and spin orientations projected onto the unit sphere.

The sequence of phases reflects the competition between exchange, DM interactions, and Zeeman energy: the helical state (Fig.~\ref{fig:SnapPhases}(a)) minimizes exchange and DM energies at low field, while increasing $B$ favors spin alignment, leading first to the triple-$\qv$ skyrmion crystal state, going through the skyrmion gas, where skyrmions persist without long-range order, (Figs.~\ref{fig:SnapPhases}(b) and (c)) and eventually to the fully polarized phase (Fig.~\ref{fig:SnapPhases}(d)). This sequence is consistent with the standard picture in chiral magnets, where the DM interaction selects a finite modulation length scale, while the magnetic field controls the density of topological defects.

In addition to these regimes, a skyrmion-bimeron (SkX-Bm) phase appears as the magnetic field increases from the helical phase: helices break into bimerons (or elongated skyrmions) \cite{leonov2024}. This can be understood as a partial unwinding of the helical modulation, where the competition between Zeeman and chiral interactions favors localized topological objects instead of extended spirals. Such transformations are often associated with a tendency toward in-plane spin configurations, where skyrmions continuously deform into elongated or split textures, as reported in systems with competing anisotropies. This interpretation differs from the alternative definition based on bound meron pairs~\cite{Gobel2019}. 

%%%
\subsubsection{Square double-$\qv$ skyrmion ($2\qv$-SkX) phase - ESP and BP models}
%%%
A representative configuration is shown in Fig.~\ref{fig:SnapPhases}(e) for $\{B/J_0,\alpha,\beta\} = \{0.4, 0.1, 0.7\}$ in the ESP model. It corresponds to an approximate square skyrmion lattice with four-fold symmetry. The structure factor in this phase shows symmetric peaks in orthogonal wave vectors $\qv_1$ and $\qv_2$, reflecting its double-$\qv$ character and four-fold rotational symmetry. Secondary peaks in orthogonal directions are also seen. A similar pattern for a skyrmion square lattice was presented in Ref.~[\onlinecite{hayami2021square}], where the square skyrmion lattice was stabilized with effective bilinear and biquadratic interactions from an effective spin model arising from a  Kondo lattice model considering itinerant electrons and localized spins. Experimentally, a compact square skyrmion lattice was measured and visualized through Lorentz transmission electron microscopy in the centrosymmetric tetragonal magnet GdRu$_2$Si$_2$ \cite{khanh2020squareexperiment}. Here, we refer to this low-temperature phase as an ``approximate'' square skyrmion lattice since the skyrmions are not arranged in a perfect square arrangement, which is reflected in the emergence of additional third-order peaks in reciprocal space. 

This phase can be interpreted as arising from the competition between interactions that favor different modulation directions, leading to the coexistence of two orthogonal ordering wave vectors instead of the single-$\qv$ or triple-$\qv$ states typical of helical and hexagonal skyrmion phases. In contrast to the conventional hexagonal SkX, this square arrangement indicates a modification of the effective interaction landscape, where the degeneracy between different ordering directions is lifted, favoring orthogonal modulations. This type of phase, as discussed above, is also stabilized in the BP model at larger magnetic fields. Given that for both SP coupling models the phase arises at large $\beta$ and negligible $\alpha$, our results suggest that the competition between the linear and quadratic DM terms drives the stabilization of this type of phase. In the next subsection we will present further arguments to support this, and discuss the differences in both models: in the ESP case, the 2$\qv$-SkX phase comes from a triple-$\qv$ as $\beta$ increases, whereas in the BP model the 2$\qv$-SkX textures are stabilized from a ferromagnetic order.

%%%
\subsubsection{Skyrmion-bimeron (SkX-Bm) phase - ESP and BP models}
%%%

Fig.~\ref{fig:SnapPhases}(f) shows a configuration with coexisting skyrmions and bimerons. Both carry the same topological charge \cite{ezawa2011compact,silva2014emergence,rosales2023skyrmion,gomez2024chiral}. Bimerons are known to emerge in the presence of local fields, disorder, and thermal fluctuations\cite{silva2014emergence,iroulart2024probing,ezawa2011compact}. This phase appears in both models, but is more prominent in the BP case. It reflects a regime where different length scales compete, allowing localized skyrmions and elongated bimerons to coexist without forming a single periodic structure. This coexistence suggests that the system is close to a crossover regime between different topological textures, where small changes in parameters can favor either compact skyrmions or elongated bimeron-like structures. In fact, this can be seen in the BP model, where for smaller magnetic fields (see Fig.~\ref{fig:GridSnapsBP}(b)), at larger $\alpha$ and small $\beta$, the SkX-Bm phase appears as an intermediate phase between helices and skyrmions induced by larger $\alpha$, which in the BP model is associated with a quadratic exchange term that favors the parallel alignment of neighbouring spins. This favouring, combined with the ferromagnetic exchange interaction, acts similarly to an external magnetic field. This is different in the ESP model: bimerons arise from spirals as $\beta$ is increased, for low $\alpha$ (Fig.~\ref{fig:GridSnapsESP}(b)):  the quadratic DM term seems to favor chiral arrangements (without fixing the helicity), but the effective three-site DM term is incompatible with the in-plane circulation on skyrmions, thus favoring the elimination of these textures for large enough $\beta$.

%%%
\subsubsection{Triple-$\qv$ Skyrmion-crystal (SkX) phase - ESP and BP models}
%%%
As discussed before, the additional interactions arising from the SP coupling models used in this work enhance and give rise to the well-known triple-$\qv$ skyrmion crystal phase, where skyrmions form a periodic arrangement. In Figs.~\ref{fig:SnapPhases}(j),(l) we present two examples of this type of phase induced by $\alpha$ and $\beta$ in the BP model for magnetic fields, where no skyrmions are found in the non-phonon interacting case. At lower fields, the SkX phase is stabilized for large $\alpha$, following the behavior previously disregarded for bimerons. At higher fields, however, the picture changes: coming from the polarized phase, the competition between exchange and DM interactions is reinforced by the quadratic DM and exchange terms in Eq.(\ref{HamiltonianEff2}), inducing SkX in a broad region in the $(\alpha,\beta)$ parameter space.

%%%
\subsubsection{Distorted helices and labyrinth phases - ESP and BP models}

For both SP coupling schemes, at lower fields and low $\alpha$, increasing $\beta$ leads to the distortion of helices due to the increasing weight of terms that favor the perpendicular alignment of spins, destroying perfect helices and leading to labyrinth-like phases, as illustrated in Fig.~\ref{fig:SnapPhases}(i).
%%%

%%%

\subsubsection{Meron phases -  ESP model}
%%%

A particular feature of the ESP model is the emergence of periodic arrangements of merons and antimerons at low magnetic fields, strong $\alpha$ and small $\beta$.  Merons \cite{lin2015skyrmion}, which have half the topological charge of a skyrmion, are expected to exist, for example in centrosymmetric magnets. It has been shown that competing easy-axis and compass anisotropies, producing a relatively weak effective easy-plane anisotropy, stabilize meron crystals formed by an alternating pattern of core-up and core-down merons \cite{wang2021meron}. Moreover, recent experiments have shown a transformation between meron and skyrmion topological spin textures in the chiral magnet Co$_8$Zn$_9$Mn$_3$ \cite{yu2018transformation}. Additionally, meron crystals have been shown to arise from spiral spin liquids in triangular-based lattices \cite{mohylna2025frustration}, such as seen experimentally in Gd$_3$Ru$_4$Al$_{12}$ \cite{mo2025skyrmionmeron}.

In these aforementioned cases, the magnetic unit cell hosts a combination of core-up and core-down merons and antimerons in such a way that there is a net topological charge.  However, as shown in detail in Fig. \ref{fig:SnapPhases}(g), the meron-antimeron lattice that arises in the ESP model is formed by a periodic arrangement of bottom-down merons and bottom-up antimerons, formed by $4\times 4 $ spin clusters. This corresponds in reciprocal space to a double-$\qv$ structure with orthogonal helical $\qv$ vectors, resulting in four Bragg peaks, as illustrated in the inset of Fig. \ref{fig:SnapPhases}(g). Given that there are the same number of merons and antimerons, the phase has a zero total topological charge, which is reflected in the scalar chirality, as we discuss in the following subsection.

In the larger-$\alpha$ and small-$\beta$ regime, the system forms a more compact variant of the meron–antimeron crystal, which we refer to as a compact meron–antimeron crystal phase (CM-aM), presented in Fig.~\ref{fig:SnapPhases}(h). It consists of localized $2\times 2$ regions of nearly aligned spins with an in-plane circulation, and exhibits four Bragg peaks at larger wave vectors, indicating a shorter periodicity.
This reduction in length scale suggests that the effective interactions favor more localized topological defects, leading to a higher density of meron–antimeron structures.

This behaviour indicates the dominance of the three-site $\alpha$ term in the effective Hamiltonian of Eq.~(\ref{HamiltonianEff1}), which favors an order where each spin is parallel to one neighbor and antiparallel to the other neighbor in the opposite direction. In a pure exchange model, this would induce a periodic tiling of $2\times 2$ clusters ordered antiferromagnetically, competing with the Zeeman term and lowering the net magnetization. The DM interaction induces a non-zero chirality in each cluster, giving rise to merons and antimerons.  
\begin{figure}[h!]
\includegraphics{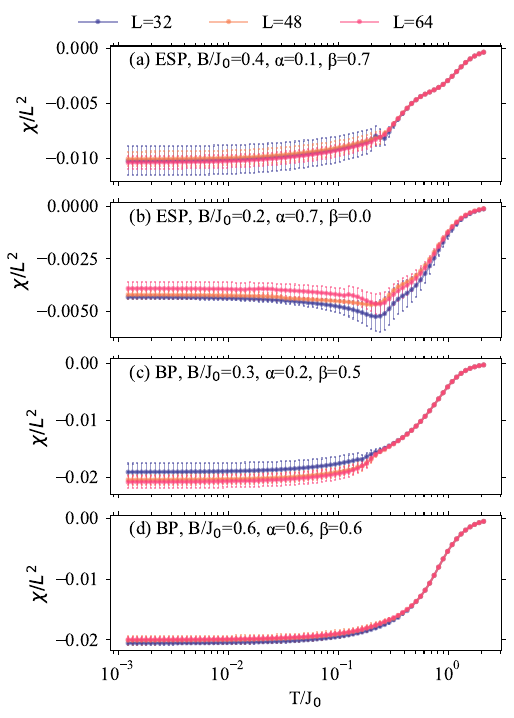}
\caption{Temperature dependence of the normalized chirality $\chi/L^2$ for different system sizes $L=32$, $48$, and $64$ (shown in blue, orange, and pink, respectively) for the Einstein site-phonon and bond-phonon coupling schemes, averaged over 15 independent copies. (a) ESP model in $2\qv$-SkX phase ($B/J_0=0.4$, $\alpha=0.1$, $\beta=0.7$),(b) ESP model in M-aM phase ($B/J_0=0.2$, $\alpha=0.7$, $\beta=0.0$), (c) BP model in SkX-Bm phase ($B/J_0=0.3$, $\alpha=0.2$, $\beta=0.5$), and (d) BP model in SkX phase ($B/J_0=0.6$, $\alpha=\beta=0.6$). The scaling reveals distinct temperature dependencies for each phase, with $\chi/L^2$ serving as an intensive measure of chiral order.}
\label{chi_scaled}
\end{figure}
\begin{figure*}[th!]
\includegraphics[width=\textwidth]{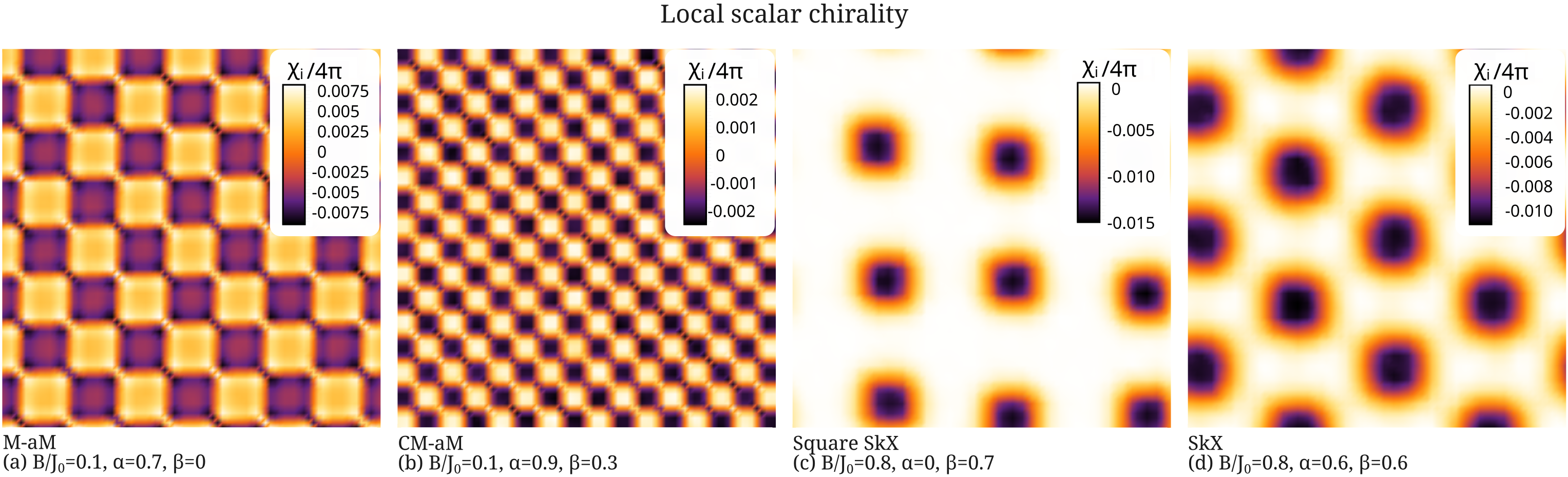}
\caption{Local scalar chirality density $\chi_i$ for representative phases discussed in Fig.~\ref{fig:SnapPhases}: (a) Meron-antimeron crystal (M-aM), (b) Compact Meron-antiMeron crystal (CM-aM), (c) Square double-$\qv$ skyrmion crystal ($2\qv$-SkX), and (d) skyrmion crystal (SkX). Positive and negative values correspond to opposite local chiralities.}
\label{fig:top_density}
\end{figure*}

\subsection{Order parameters}

As was established in the previous subsections, a surprising variety of topological low temperature phases emerge in both SP schemes in different ranges of the effective couplings $\alpha$ and $\beta$. In this subsection, we focus in detail in order parameters to 
explore them. First, we further focus into the scalar chirality, the typical order parameter for topological phases used to construct the phase diagrams from Fig.~\ref{fig:PDChir}. Secondly, we introduce order parameters to distinguish the square skyrmion crystal  double-$\qv$ phase from the well-known hexagonal  triple-$\qv$ skyrmion crystal.

\subsubsection{Scalar chirality}

We have previously shown phase diagrams, built as density plots of the absolute value of the scalar normalized chirality. As discussed above, a plethora of topological phases arise. Among them, the M-aM crystals stabilized in this work are expected to have $\chi \sim 0$. In this subsection, we focus on two objectives: first, we analyze the stability of the phases with system size by studying the scalar normalized chirality. Secondly, we construct an order parameter related to the chirality to analyze the emergence of the M-aM crystals and show their topological nature.

For our first objective, we choose four different phases, two in each phonon-site model: the $2\qv$-SkX phase ($B/J_0=0.4$, $\alpha=0.1$, $\beta=0.7$) and the M-aM crystal ($B/J_0=0.2$, $\alpha=0.7$, $\beta=0.0$) for the ESP model; the hexagonal SkX phase ($B/J_0=0.3$, $\alpha=0.2$, $\beta=0.5$) and the SkX-Bm phase ($B/J_0=0.6$, $\alpha=\beta=0.6$) for the BP case. In Fig.~\ref{chi_scaled}, we plot $\chi/L^2$ for three system sizes, $L=32,48,64$, averaged over $15$ independent copies. Results show that indeed the phases are stable with system size, supporting their topological nature, with the exception of the M-aM phase, where $\chi \sim 0$ at low temperatures.

 The M-aM crystal found in this work is a periodic arrangement of pairs of merons and antimerons. Since merons and antimerons have opposite topological charge and their sizes in these textures are the same, the total scalar chirality of the system is going to be close to zero, as shown before. Thus, we resort to other strategies to explore these phases and show their topological nature. First, we study the distribution of the topological charge in different phases. In a system where skyrmions are found, we expect to find a core with non-trivial topological charge, corresponding to the location of the skyrmion, and a zero topological charge background. For a M-aM crystal, a pattern alternating positive and negative values of the topological charge is expected, leading to a total value of  close to zero. To visualize this cancellation mechanism, in Fig.~\ref{fig:top_density} we show the scalar chirality density $\chi_{i}$, defined as the triple product of the triangles defining the scalar chirality at site $i$, for representative phases. 
Panels (a) and (b) correspond to the crystals M-aM and CM-aM, where positive and negative chirality regions alternate periodically, leading to a nearly vanishing net scalar chirality despite the presence of locally nontrivial textures. Panels (c) and (d) illustrate a double$-\qv$ square and triple-$\qv$ hexagonal skyrmion lattices, respectively. In these skyrmion phases, the system exhibits chirality of the same sign across the system, resulting in a finite net value. In these cases, the presence of skyrmions is easily identified with a negative chirality region, surrounded by a non-chiral region, connected with the polarized background. Comparing panels (a-b) with (c-d), it can be easily appreciated that, while the four textures are topologically non-trivial, the scalar chirality in (a-b) will add up to zero, whereas in (c-d) the textures have a net scalar chirality.

\begin{figure}[h!]
\includegraphics{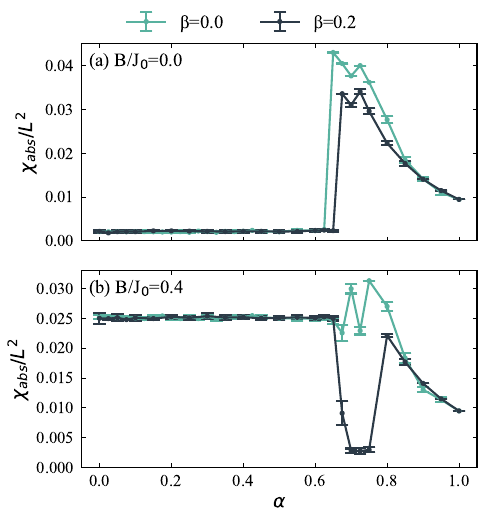}
\caption{Total absolute chirality $\chi_{abs}/L^2$ for $B/J_0=0$ (panel (a)) and $B/J_0=0.4$ (panel (b)) for the Einstein site-phonon model as a function of $\alpha$ for fixed $\beta=0.0$ and $\beta=0.2$ and at the lowest simulated temperature $T/J_0 = 1.2 \times 10^{-3}$.}
\label{abschi_alpha}
\end{figure}

Furthermore, considering the discussion above, to further study the meron-antimeron phases, for the ESP model we compute what we define as the total absolute chirality $\chi_{abs} = \frac{1}{4\pi}  \left\langle \sum_{i}|\Sv_{i} \cdot (\Sv_{j} \times \Sv_{m})| + |\Sv_i\cdot(\Sv_{k} \times \Sv_{n})| \right\rangle$, calculated as the sum of the absolute value of the chirality on each lattice triangle, aiming to prevent cancellation of the total sum.  A similar approach was used for fractional antiferromagnetic skyrmions \cite{gao2020fractional}. Fig.~\ref{abschi_alpha} shows $\chi_{abs}$ as a function of $\alpha$ for fixed $\beta=0.0$ and $\beta=0.2$ for two values of the external magnetic field. For $B/J_0=0$, a peak around $\alpha\approx0.7$ signals the transition from the helical phase to the M-aM phase, stabilized at low magnetic fields for large $\alpha$ and low $\beta$. This abrupt change is not visible in the net chirality, due to cancellation between opposite contributions from merons and antimerons, but is clearly captured by $\chi_{abs}$, which gets lower as $\alpha$ increases since it favors an arrangement of parallel and antiparallel spin clusters. For $B/J_0=0.4$, the low-$\alpha$ region corresponds to the SkX phase, which exhibits a larger $\chi_{abs}$ than the intermediate M-aM regime. A noticeable feature in the $\beta=0.2$ curve is the abrupt drop at $\alpha\sim 0.6$, which corresponds to an almost canted spiral phase, an intermediate phase between the SkX and the M-aM phases. Due to the presence of chiral domains, the curve does not drop to zero, but $\chi_{abs}$ gets significantly lower in this region, before going up at $\alpha\sim0.8$, into the compact M-aM crystal.

\subsubsection{Skyrmion coordination number}

\begin{figure}[t]
    \centering
    \includegraphics[width=\linewidth]{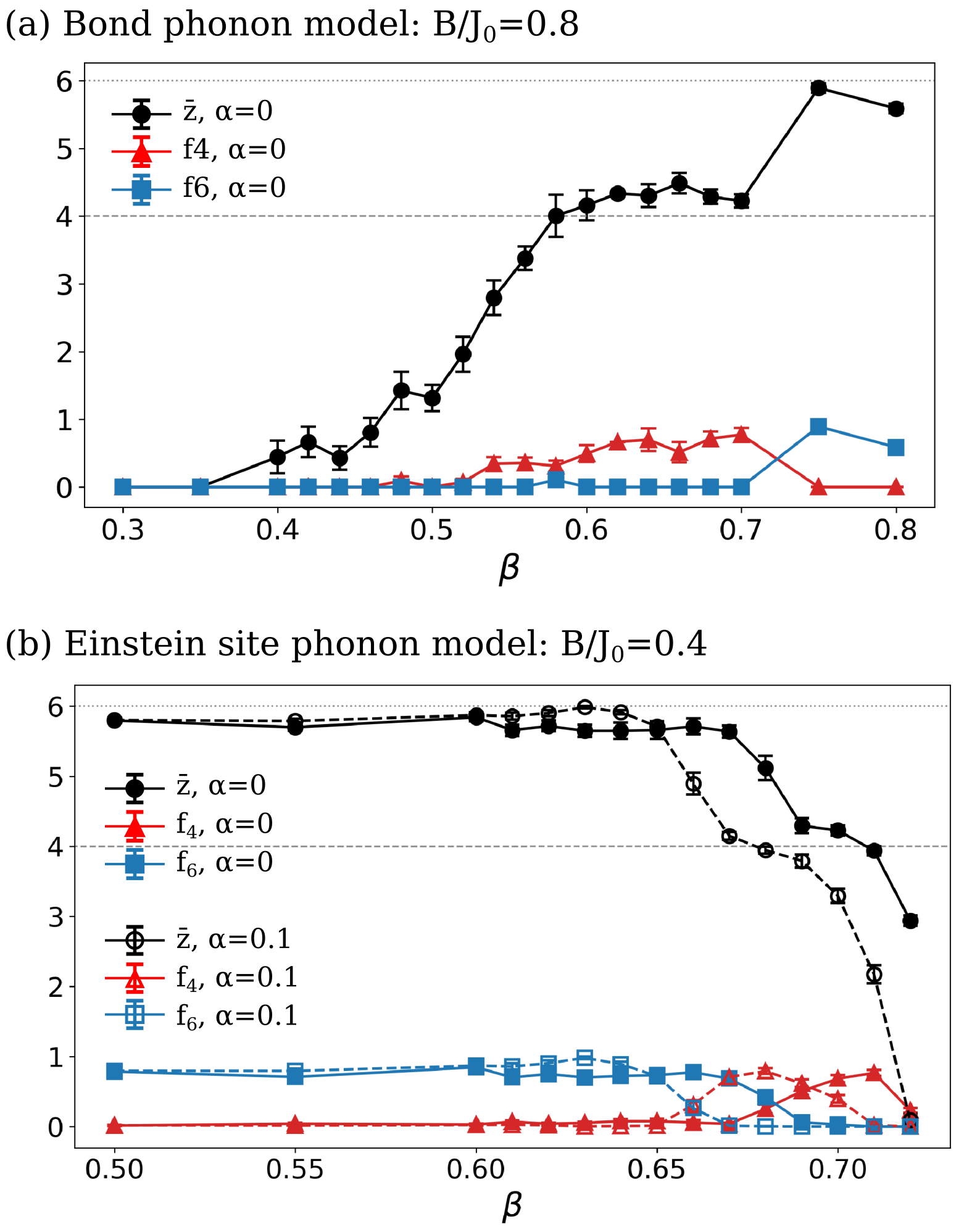}
    \caption{
    Real-space coordination analysis of the skyrmion-core lattice at the lowest simulated temperature $T/J_0 = 1.2 \times 10^{-3}$.
    The average coordination number $\bar z$ (black circles) and the fractions of fourfold and sixfold coordinated skyrmion cores, $f_4$ (blue triangles)  and $f_6$ (red squares), are shown as functions of $\beta$ . 
    (a) Bond-phonon model at $B/J_0=0.8$ and $\alpha=0$.
    (b) Einstein site-phonon model at $B/J_0=0.4$, comparing $\alpha=0$ and $\alpha=0.1$ (filled and empty markers, solid and dashed lines, respectively).
    The horizontal lines mark the ideal square- and triangular-lattice coordination values, $z=4$ and $z=6$, respectively. 
    }
    \label{fig:coordination_skx}
\end{figure}

We have previously shown that, in both SP coupling effective models, a double-$\qv$ skyrmion phase is stabilized at low temperatures with $\beta$ for small $\alpha$. In the ESP model,   the system goes from the triple-$\qv$ triangular (hexagonal) skyrmion lattice to the  double-$\qv$ square skyrmion phase (see Fig.~\ref{fig:GridSnapsESP}), whereas in the BP model the square lattice emerges with larger $\beta$ from the ferromagnetic phase. In both cases, the square arrangement of skyrmions is not a perfect square lattice, but inspection of the structure factor and the real space textures shows that it is clearly a different phase from the triple-$\qv$ lattice and also from a skyrmion gas, a texture where skyrmions do not form an ordered pattern. However, traditional order parameters such as the total scalar chirality of the lattice are not sufficient to distinguish between these two arrangements.  A key difference between the square double-$\qv$ and the hexagonal triple-$\qv$ skyrmion lattices is the coordination number of each skyrmion: in a square lattice each skyrmion has 4 nearest neighbors, whereas there are 6 in a hexagonal tiling.   Thus, we resort to the skyrmion coordination number as an order parameter. For each low-temperature snapshot, the skyrmion centers were identified from the local minima of $S^z_i$, and a nearest-neighbor lattice was constructed taking into account that the simulations were done using periodic boundary conditions. From this skyrmion center lattice, we computed the average coordination number
\begin{equation}
    \bar z = \frac{1}{N_{\rm sk}}\sum_{i=1}^{N_{\rm sk}} z_i ,
\end{equation}
together with the fractions
\begin{equation}
    f_n = \frac{1}{N_{\rm sk}}\sum_{i=1}^{N_{\rm sk}} \delta_{z_i,n},
    \qquad n=4,6 ,
\end{equation}
where $N_{\rm sk}$ is the number of detected skyrmion cores and $z_i$ is the coordination of the $i$-th core. In this way, $f_4$ provides a direct real-space indicator of a square-like skyrmion arrangement, while $f_6$ signals a triangular-like skyrmion lattice. The results are shown in Fig.~\ref{fig:coordination_skx}. For the BP model (panel (a)) at $B/J_0=0.8$ and $\alpha=0$, the system evolves from a polarized state at small $\beta$ to a dilute skyrmion gas regime, followed by an intermediate region where $\bar z$ approaches the square-lattice value $z=4$. At larger $\beta$, the coordination increases toward values closer to those expected for a triangular skyrmion lattice. In the ESP model at $B/J_0=0.4$ (panel (b)), the coordination analysis shows that the low-$\beta$ skyrmion crystal is predominantly triangular-like, while increasing $\beta$ reduces the sixfold coordination and enhances the fourfold component. This crossover occurs at slightly lower $\beta$ for $\alpha=0.1$, consistent with Fig.~\ref{fig:GridSnapsESP}. Further increasing $\beta$ rapidly reduces the coordination number, indicating that the system is driven into a skyrmion gas phase.

To further clarify this picture, for both models in Fig.~\ref{fig:coordination_skx} we plot the fraction of skyrmion cores with coordination number 4 (red squares) and with coordination number 6 (blue triangles). For the BP model, these parameters are both close to zero for low $\beta$, consistent with a skyrmion gas phase. Then, at $\beta\sim 0.6$, the $f_4$ goes up and gets close to 1, corresponding to the onset of the square lattice, and then at higher $\beta$, $f_4$ goes to zero as $f_6$ increases, corresponding to the stabilization of the triangular lattice. In the ESP model, on the other hand, $f_6$ is close to 1 for smaller $\beta$, then goes to zero while $f_4$ rises, and for larger $\beta$ both quantities drop, signaling the onset of the skyrmion gas.
\subsection{Comparison between spin-phonon coupling models}

As we have shown throughout this work, the two spin-phonon coupling models lead to qualitatively different mechanisms for stabilizing magnetic textures. The  ESP model, where each site has an independent deformation,  exhibits a wider variety of phases, including $2\qv$-SkX, M-aM, and CM-aM states, and shows extended regions where the conventional hexagonal skyrmion lattice is suppressed. This behaviour can be understood from the effective Hamiltonian in Eq.~(\ref{HamiltonianEff1}), where multi-spin terms introduce competing interactions. In particular, the three-site $\alpha^2$ contributions tend to favor locally antiparallel configurations, while the three-site  $\beta^2$ terms promote chiral arrangements incompatible with a defined in-plane circulation as that favored by the DM interaction, leading to a strong competition that results in multiple distinct phases.

In the limit $\alpha=0$ and finite $\beta$, the system is dominated by the chiral terms, which compete with the magnetic field. This produces a sequence of textures ranging from distorted spirals at zero field to skyrmion-related phases at lower and intermediate fields. In fact, an interesting feature found here is that a large $\beta$ can distort the well-known hexagonal triple-$\qv$ skyrmion lattice into an almost square double-$\qv$ skyrmion arrangement. On the other hand, for $\beta=0$ and finite $\alpha$, the system is governed by the $\alpha^2$ terms, which favor collinearity and an alternating ferromagnetic and antiferromagnetic local order, leading to phases such as the M-aM and CM-aM crystals for large enough $\alpha$.

\begin{figure}[t]
\centering
\includegraphics[width=0.85\columnwidth]{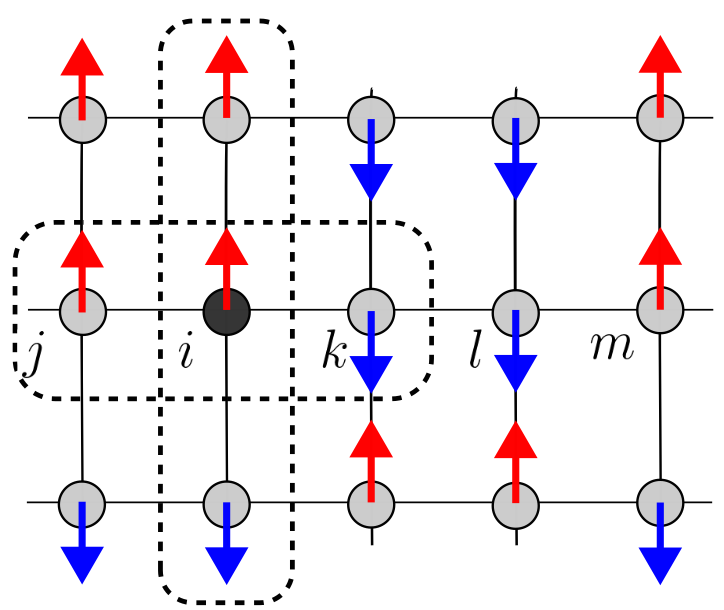}
\caption{
Local spin configurations illustrating the effect of the three-site
$\alpha^2$ interaction in the ESP model. For two opposite bonds,
$\hat{\mathbf e}_{ij}\!\cdot\!\hat{\mathbf e}_{ik}=-1$, the corresponding
contribution is proportional to
$+(\mathbf S_i\!\cdot\!\mathbf S_j)
(\mathbf S_i\!\cdot\!\mathbf S_k)$.
Configurations with opposite signs of the two scalar products lower the energy, as shown for site $i$ and its nearest neighbors. The interaction therefore promotes locally competing ferromagnetic and antiferromagnetic correlations, favoring a $2\times 2$ antiferromagnetic tiling.
}
\label{fig:alpha3site}
\end{figure}

This tendency can be understood directly from the three-site
$\alpha^2$ interaction. Consider two opposite bonds $ij$ and $ik$,
for which $\hat{\mathbf e}_{ij}\cdot\hat{\mathbf e}_{ik}=-1$.
The corresponding three-site contribution is then proportional to
\begin{equation}
+\alpha^2
(\mathbf S_i\cdot\mathbf S_j)
(\mathbf S_i\cdot\mathbf S_k).
\label{eq:alpha3site}
\end{equation}
Its energy is minimized when the two scalar products have opposite signs, i.e., when the central spin is locally parallel to one neighboring spin and antiparallel to the other. Figure~\ref{fig:alpha3site} illustrates the locally favored and unfavored spin configurations associated with this interaction. The three-site term therefore naturally promotes competing ferromagnetic and antiferromagnetic local correlations. In fact, the three-site term alone is minimized by a period-two arrangement of locally competing ferromagnetic and antiferromagnetic correlations. Combined with the DM interaction and the magnetic field, this competition provides a simple microscopic interpretation for the stabilization of the meron-antimeron textures found in the ESP model.

In contrast, the BP model retains a simpler structure, with only on-bond interactions. In the effective Hamiltonian from Eq.~(\ref{HamiltonianEff2}), this implies that the effective $\alpha^2$ and $\beta^2$ terms are only quadratic exchange and DM bond interactions, without the three-site couplings from the ESP case. As a result, the competition between terms is less frustrated, and the model more directly favors skyrmion formation. In this case, increasing $\beta$ enhances chiral interactions and stabilizes skyrmion textures over a wide range of magnetic fields, generating SkX phases even from ferromagnetic textures. In particular, at higher fields and $\alpha=0$, the square double-$\qv$ skyrmion lattice is found, supporting the idea that the quadratic DM term favors such ordering. However, this is different from the ESP case, where the double-$\qv$ skyrmion phase was found when increasing $\beta$ from the triple-$\qv$ region, and not from the ferromagnetic phase. To understand this at a microscopic level, we will resort to the low-energy continuum theory, as shown in the next subsection.

On the other hand, the role of $\alpha$ is different: it modifies the effective exchange interaction and shifts the stability region of the skyrmion phase, but does not introduce the same level of competing multi-spin effects as in the ESP model. Combined with the ferromagnetic exchange coupling, the $\alpha^2$ term acts as an external magnetic field, and skyrmions emerge at lower magnetic fields, where helices are found in the typical skyrmion Hamiltonian. Overall, the BP model exhibits smoother transitions and a more robust skyrmion region.

\subsubsection{Low-energy interpretation of the square skyrmion phases}
\label{sec:lowenergy}

To clarify the different mechanisms responsible for the stabilization of the square skyrmion crystal, we consider the continuum limit of the phonon-induced $\beta^2$ interactions. The complete derivation and additional details are presented in Appendix~\ref{appendixA}.

Assuming slowly varying spin textures, the spin field can be expanded as
\begin{equation}
\mathbf{S}(\mathbf{r}+a\hat{\mu})
=
\mathbf{n}(\mathbf{r})
+
a\,\partial_\mu\mathbf{n}
+
\frac{a^2}{2}\partial_\mu^2\mathbf{n}
+\cdots ,
\label{Eq:continuum_expansion_main}
\end{equation}
where $\mathbf n(\mathbf r)$ is a unit vector, $a$ is the lattice spacing, and $\mu=x,y$. The conventional exchange, Dzyaloshinskii-Moriya, and Zeeman interactions lead to the standard continuum functional \cite{han2010skyrmion}
\begin{equation}
\mathcal F_0
=
\int d^2r
\left[
\frac{A}{2}(\nabla\mathbf n)^2
+
D\,\mathbf n\cdot(\nabla\times\mathbf n)
-
Bn_z
\right].
\label{Eq:F0_main}
\end{equation}
The phonon-induced terms generate additional nonlinear gradient contributions. For the BP model, the leading $\beta^2$ interaction becomes
\begin{eqnarray}
\mathcal F_{\beta}^{\rm BP}
&=&
-K_{\rm BP}\int d^2r
\left[
\left(
n_y\partial_xn_z-n_z\partial_xn_y
\right)^2
\right.
\nonumber\\
&&\left.
+
\left(
n_z\partial_yn_x-n_x\partial_yn_z
\right)^2
\right],
\label{Eq:FBP_main}
\end{eqnarray}
where $K_{\rm BP}>0$ is proportional to $\beta^2D^2/\kappa$. In contrast, the ESP model produces the higher-gradient contribution
\begin{eqnarray}
\mathcal F_{\beta}^{\rm ESP}
&=&
-K_{\rm ESP}\int d^2r
\left[
\left(
n_y\partial_x^2n_z-n_z\partial_x^2n_y
\right)^2
\right.
\nonumber\\
&&\left.
+
\left(
n_z\partial_y^2n_x-n_x\partial_y^2n_z
\right)^2
\right],
\label{Eq:FESP_main}
\end{eqnarray}
with $K_{\rm ESP}>0$ proportional to $\beta^2D^2a^2/\kappa$. Thus, although both interactions originate from the quadratic DMI-related terms, they enter the long-wavelength theory at different orders in spatial derivatives.

This difference becomes clearer by expanding around the field-polarized state,
\begin{equation}
\mathbf n(\mathbf r)
=
\left(
\pi_x,\,
\pi_y,\,
\sqrt{1-\pi_x^2-\pi_y^2}
\right),
\qquad
|\pi_x|,|\pi_y|\ll1 .
\end{equation}
Keeping quadratic contributions in the transverse fluctuations and introducing their Fourier components, one obtains
\begin{equation}
\mathcal F_{\beta}^{\rm BP}
\simeq
-K_{\rm BP}
\sum_{\mathbf q}
\left[
q_x^2|\pi_y(\mathbf q)|^2
+
q_y^2|\pi_x(\mathbf q)|^2
\right],
\label{Eq:FBP_main_fourier}
\end{equation}
whereas
\begin{equation}
\mathcal F_{\beta}^{\rm ESP}
\simeq
-K_{\rm ESP}
\sum_{\mathbf q}
\left[
q_x^4|\pi_y(\mathbf q)|^2
+
q_y^4|\pi_x(\mathbf q)|^2
\right].
\label{Eq:FESP_main_fourier}
\end{equation}
Eq. ~(\ref{Eq:FBP_main_fourier}) shows that the BP interaction modifies the quadratic fluctuation spectrum already at order $q^2$. In particular, it lowers the energy of two symmetry-related axial modes: one propagating along $x$ and polarized along $y$, and the other propagating along $y$ and polarized along $x$. These modes provide the natural building blocks of a double-$\mathbf q$ square skyrmion crystal. This gives a simple interpretation of why, in the BP phase diagram, increasing $\beta$ can destabilize the polarized phase directly toward a square skyrmion state (see Fig.~\ref{fig:GridSnapsBP}, panel (d)).

The ESP interaction behaves differently. Its contribution appears only at order $q^4$ and therefore does not modify the leading long-wavelength instability of the polarized phase. This agrees with the numerical results, where the square skyrmion crystal does not emerge directly from the polarized state, but instead appears through a reconstruction of an already established triangular skyrmion crystal (see Fig.~\ref{fig:GridSnapsESP}, panels (c) and (d)).

To obtain qualitative insight into this second mechanism related to the ESP model, we compare simple multi-$\mathbf q$ variational textures. We introduce the auxiliary magnetization field
\begin{eqnarray}
\mathbf n(\mathbf r)
&=&
\left[
n_0+
A_z\sum_{\nu=1}^{N_q}
\cos(\mathbf Q_\nu\cdot\mathbf r+\phi_\nu)
\right]\hat{\mathbf z}
\nonumber\\
&&+
A_\perp\sum_{\nu=1}^{N_q}
\left(
\hat{\mathbf z}\times\hat{\mathbf Q}_\nu
\right)
\sin(\mathbf Q_\nu\cdot\mathbf r+\phi_\nu),
\label{Eq:multiq_ansatz_main}
\end{eqnarray}
where $n_0$, $A_z$, and $A_\perp$ are variational amplitudes
controlling the uniform, longitudinal, and transverse components,
respectively, while $\phi_\nu$ denotes the phase of each mode.
For this simple energetic comparison, the local hard-spin constraint
$|\mathbf n(\mathbf r)|=1$ is not imposed. The ansatz is therefore used
only to identify the energetic tendency of the phonon-induced
interaction and not as a complete variational solution of the
hard-spin model. The square state is represented by two orthogonal modes,
\begin{equation}
\mathbf Q_1=(Q,0),
\qquad
\mathbf Q_2=(0,Q),
\end{equation}
whereas the triangular state contains three modes separated by $120^\circ$,
\begin{eqnarray}
\mathbf Q_1&=&Q(1,0),\nonumber\\
\mathbf Q_2&=&Q\left(-\frac{1}{2},\frac{\sqrt{3}}{2}\right),\nonumber\\
\mathbf Q_3&=&Q\left(-\frac{1}{2},-\frac{\sqrt{3}}{2}\right),
\end{eqnarray}
with $\mathbf Q_1+\mathbf Q_2+\mathbf Q_3=0$.

Evaluating only the additional ESP contribution in Eq.~(\ref{Eq:FESP_main}) and averaging over one magnetic unit cell gives
\begin{equation}
\overline{\mathcal F}_{\beta,\square}^{\rm ESP}
=
-K_{\rm ESP}A_\perp^{\,2}Q^4
\left(
n_0^2+\frac{1}{2}A_z^2
\right)
\label{Eq:EsquareESP_main}
\end{equation}
for the square state, while for the triangular state one obtains
\begin{equation}
\overline{\mathcal F}_{\beta,\triangle}^{\rm ESP}
=
-K_{\rm ESP}A_\perp^{\,2}Q^4
\left[
\frac{15}{16}n_0^2
+
\frac{9}{16}A_z^2
+
\frac{9}{16}n_0A_z\cos\Phi
\right],
\label{Eq:EtriangleESP_main}
\end{equation}
where $\Phi=\phi_1+\phi_2+\phi_3$. For the usual skyrmion phase relation $\Phi=\pi$, the difference between the corresponding energy gains is
\begin{equation}
G_{\square}^{\rm ESP}
-
G_{\triangle}^{\rm ESP}
=
\frac{1}{16}
\left(
n_0^2-A_z^2+9n_0A_z
\right),
\label{Eq:ESP_energy_difference_main}
\end{equation}
where
$\overline{\mathcal F}_{\beta}^{\rm ESP}
=-K_{\rm ESP}A_\perp^{\,2}Q^4G^{\rm ESP}$.
For a finite positive uniform magnetization and comparable values of $n_0$ and $A_z$, Eq.~(\ref{Eq:ESP_energy_difference_main}) is positive. Within this simplified ansatz, the $\beta^2$ interaction can therefore lower the energy of the square two-$\mathbf q$ texture more strongly than that of the triangular three-$\mathbf q$ state.

This variational comparison does not constitute a complete determination of the triangular-to-square phase boundary. The full energy balance also includes exchange, bilinear Dzyaloshinskii-Moriya interaction, magnetic field, higher harmonics, and the local hard-spin constraint. Nevertheless, it illustrates that the two phonon models can stabilize square skyrmion crystals through different mechanisms: the BP interaction modifies the leading instability of the polarized phase, whereas the ESP interaction mainly changes the nonlinear energetic competition between already established multi-$\mathbf q$ textures.

%
%%%%%%%%%%%%%%%%%%%%%
\section{Discussion and Conclusions}
\label{sec:conclusions}

The main goal of this work was to study the effect of spin-phonon coupling on two-dimensional skyrmion systems, focusing on how lattice distortions modify magnetic interactions and distort, expand, and induce spin textures, focusing on topological low temperature phases. We proposed two models for the lattice deformation, leading to two effective Hamiltonians: the ESP model, where each site deformation is treated independently, and the BP model, where each nearest-neighbor link is assigned a displacement. Starting from a typical skyrmion Hamiltonian in the square lattice, the emergent effective interactions are governed by parameters $\alpha$ and $\beta$, associated with the exchange and DM interaction deformations, respectively. Using Monte Carlo simulations, for each model, we explored the parameter space defined by $\alpha$ and $\beta$ under an external magnetic field, and characterized the corresponding  low-temperature magnetic phases.  In order to do this, focusing on the emergent topological textures, we worked with the scalar chirality, a well-known order parameter related to skyrmion phases. We also defined the local chirality density and the total absolute scalar chirality, two parameters designed to capture topological textures with alternating chirality, which would otherwise be missed using the standard scalar chirality parameter.

A comparison between the two coupling schemes highlights qualitative differences. The ESP model generates effective multi-spin interactions that introduce frustration and promote a wider variety of phases, often destabilizing conventional skyrmion lattices. In this case, spin-phonon coupling leads to an extended variety of phases, with the appearance of several nontrivial magnetic textures. In particular, we identify $2\qv$-SkX lattices, mixed SkX-Bm phases, and M-aM crystals. Among these, meron–antimeron phases emerge over a broad region of parameter space and results from the competition between exchange, DM interaction, and the three-site antiferromagnetic lattice-induced interactions. In contrast, the BP model involves only on-bond corrections, and in general, its most significant result is the tendency to stabilize hexagonal skyrmion crystals, even in regions where the uncoupled system would display helical or ferromagnetic order.

It is worth emphasizing that the novelty of the present results does not rely on the individual magnetic textures themselves. Hexagonal and square skyrmion crystals, bimerons, and meron-antimeron crystals have been previously reported in different magnetic models and through different microscopic mechanisms. The main result of the present work is instead that spin-phonon coupling provides a common mechanism for stabilizing a wide variety of these textures starting from a conventional chiral-magnet Hamiltonian. Integrating out the lattice degrees of freedom generates nonlinear magnetic interactions which are qualitatively different for the ESP and BP models, leading to substantially different phase diagrams.

A particularly illustrative example is the square skyrmion crystal. In the BP model it emerges directly from the polarized state, whereas in the ESP model it results from the reconstruction of an already established triangular skyrmion lattice. As discussed in Sec.~\ref{sec:lowenergy} and Appendix~\ref{appendixA}, these two behaviors can be traced back to the different low-energy structure of the phonon-induced interactions. In addition, the three-site interactions generated in the ESP model introduce an additional source of competition that is absent in the BP model and favors the emergence of more complex magnetic textures, including meron-antimeron phases. Therefore, the comparison between the two models shows that not only the strength, but also the microscopic character of the spin-phonon coupling can play an important role in selecting topological magnetic phases.

Overall, these results show that spin-phonon coupling provides a mechanism to modify magnetic textures through lattice degrees of freedom. Depending on how the coupling is implemented, it can either favor or suppress skyrmion phases, rearrange skyrmion lattices, or stabilize alternative topological states such as M-aM crystals. This suggests that lattice distortions can be used to tune magnetic phases in low-dimensional systems where spin-phonon effects are relevant. From an experimental standpoint, our findings suggest that applying mechanical strain or pressure in chiral magnets and frustrated magnetic compounds-such as Co-Zn-Mn alloys or quasi-two-dimensional magnets-could serve as an active mechanism to tune transitions between skyrmion lattices and meron-antimeron crystals. 

Future research directions could explore the extension of these spin-phonon models to three-dimensional systems, the study of finite-temperature spin dynamics, the possibility of metastable or intermediate phases, or the integration of magnetoelastic couplings derived from first-principles calculations to quantitatively predict these topological phase diagrams in specific real materials.

%%%%%%%%%%%%%%%%%%%%%
\section*{Acknowledgments} 
%%%%%%%%%%%%%%%%%%%%%

The authors are partially supported by CONICET (PIP 2021-112200200101480CO),  SECyT UNLP PI+D  X947 and X1065, and PICT-2020 - SERIEA-03205.

\appendix
\section{Low-energy interpretation of the phonon-induced interactions}
\label{appendixA}

The Monte Carlo simulations presented above show that the Einstein site-phonon and bond-phonon models stabilize similar topological textures through qualitatively different mechanisms. In particular, while the BP model stabilizes the square skyrmion crystal directly from the polarized phase at high magnetic fields, in the ESP model the square lattice emerges through a reconstruction of an already established triangular skyrmion crystal. This naturally raises the question of how the phonon-induced interactions modify the energetic competition between different magnetic textures.

To gain further physical insight, in this section we analyze the continuum limit of the effective $\beta^2$ interactions. The purpose of this discussion is not to determine the phase boundaries analytically, which requires minimizing the full nonlinear hard-spin Hamiltonian, but rather to identify the different low-energy mechanisms generated by the two spin-phonon couplings.

\subsubsection{Continuum limit}

The continuum limit is obtained by assuming slowly varying spin textures and expanding the spin field around each lattice site according to\cite{osorio2017composite,wang2021meron}
\begin{equation}
\mathbf{S}(\mathbf{r}+a\hat{\mu})=\mathbf{n}(\mathbf{r})
+a\,\partial_\mu\mathbf{n}
+\frac{a^2}{2}\partial_\mu^2\mathbf{n}
+\cdots ,
\end{equation}
where $\mathbf{n}(\mathbf r)$ is a unit vector describing the local magnetization direction, $a$ is the lattice spacing and $\mu=x,y$. Keeping the lowest non-vanishing contributions, the conventional exchange, Dzyaloshinskii-Moriya and Zeeman terms recover the well-known continuum functional for chiral magnets

\begin{equation}
\mathcal F_0=\int d^2r\left[\frac{A}{2}(\nabla\mathbf n)^2+
D\,\mathbf n\cdot(\nabla\times\mathbf n)-Bn_z\right],
\label{Eq:F0}
\end{equation}

where $A$ and $D$ denote the effective exchange stiffness and Dzyaloshinskii-Moriya coupling, respectively.

The phonon-induced interactions generate additional nonlinear gradient terms whose form depends on the microscopic spin-phonon coupling. For the bond-phonon (BP) model, the quadratic chiral interaction gives rise to the continuum contribution

\begin{eqnarray}
\mathcal F_{\beta}^{\rm BP}&=&
-K_{\rm BP}\int d^2r\left[\left(n_y\partial_x n_z-n_z\partial_x n_y\right)^2 \right.
\nonumber\\
&&+
\left.\left(n_z\partial_y n_x-n_x\partial_y n_z\right)^2\right],
\label{Eq:FBP}
\end{eqnarray}

where
\begin{equation}
K_{\rm BP}=\frac{\beta^2D^2a^2}{2\kappa},
\end{equation}
up to an overall numerical prefactor depending on the continuum convention.

The Einstein site-phonon (ESP) model leads to a qualitatively different low-energy contribution. In this case, the lowest non-vanishing correction generated by the $\beta^2$ interaction contains second-order spatial derivatives,
\begin{eqnarray}
\mathcal F_{\beta}^{\rm ESP}&=&-K_{\rm ESP}\int d^2r\left[\left(n_y\partial_x^2 n_z-n_z\partial_x^2 n_y\right)^2 \right.\nonumber\\
&&+\left.\left(n_z\partial_y^2 n_x-n_x\partial_y^2 n_z\right)^2\right],
\label{Eq:FESP}
\end{eqnarray}

with

\begin{equation}
K_{\rm ESP}=\frac{\beta^2D^2a^4}{2\kappa},
\end{equation}
again up to an overall numerical constant.

An important difference between the two models immediately follows from Eqs.~(\ref{Eq:FBP}) and (\ref{Eq:FESP}). The bond-phonon interaction produces a contribution involving first-order gradients of the spin field, whereas the Einstein site-phonon model contributes only through second-order derivatives. As a consequence, the BP interaction directly modifies the long-wavelength energetics of spiral states, while the ESP correction becomes relevant only for more strongly modulated textures where higher-order spatial variations are important.
\subsubsection{Expansion around the polarized state}

Near the fully polarized ferromagnetic state, the magnetization can be parametrized as

\begin{equation}
\mathbf n(\mathbf r)=\left(\pi_x,\,\pi_y,\,\sqrt{1-\pi_x^2-\pi_y^2}\right),
\end{equation}
where $|\pi_x|,|\pi_y|\ll1$. To quadratic order one has

\begin{equation}
n_z\simeq1-\frac{\pi_x^2+\pi_y^2}{2},\qquad\partial_\mu n_z=\mathcal O(\pi\,\partial_\mu\pi),
\end{equation}

so that only terms linear in $\pi_{x,y}$ need to be retained.

For the bond-phonon model, Eq.~(\ref{Eq:FBP}) reduces to

\begin{equation}
\mathcal F_{\beta}^{\rm BP}\simeq-K_{\rm BP}\int d^2r\left[(\partial_x\pi_y)^2+(\partial_y\pi_x)^2\right].
\label{Eq:FBPquad}
\end{equation}

By contrast, the Einstein site-phonon contribution becomes

\begin{equation}
\mathcal F_{\beta}^{\rm ESP}\simeq-K_{\rm ESP}\int d^2r\left[(\partial_x^2\pi_y)^2+(\partial_y^2\pi_x)^2\right],
\label{Eq:FESPquad}
\end{equation}
which only contributes through fourth-order gradients.

Therefore, the two spin-phonon couplings modify the long-wavelength physics in qualitatively different ways. The BP interaction contributes already at quadratic order in both the spin fluctuations and spatial gradients, directly competing with the exchange stiffness and the conventional Dzyaloshinskii-Moriya interaction. In contrast, the ESP correction only enters through higher-order gradients and is therefore expected to play a weaker role close to the polarized state, becoming relevant only for strongly modulated magnetic textures.

\subsubsection{Fourier representation}

The different roles of the two phonon-induced interactions become even more transparent in momentum space. Introducing the Fourier components of the transverse spin fluctuations,

\begin{equation}
\pi_{\alpha}(\mathbf r)=
\frac{1}{\sqrt{N}}\sum_{\mathbf q}\pi_{\alpha}(\mathbf q)e^{i\mathbf q\cdot\mathbf r},\qquad\alpha=x,y,
\end{equation}
Eqs.~(\ref{Eq:FBPquad}) and (\ref{Eq:FESPquad}) become

\begin{equation}
\mathcal F_{\beta}^{\rm BP}
=
-K_{\rm BP}
\sum_{\mathbf q}
\left[
q_x^2|\pi_y(\mathbf q)|^2
+
q_y^2|\pi_x(\mathbf q)|^2
\right].
\label{Eq:FBPFourier}
\end{equation}
\begin{equation}
\mathcal F_{\beta}^{\rm ESP}
=
-K_{\rm ESP}
\sum_{\mathbf q}
\left[
q_x^4|\pi_y(\mathbf q)|^2
+
q_y^4|\pi_x(\mathbf q)|^2
\right].
\label{Eq:FESPFourier}
\end{equation}

The BP interaction therefore modifies the quadratic dispersion of the low-energy spin fluctuations. In particular, it lowers the energy of transverse modes propagating along the principal lattice directions, providing a simple microscopic explanation for why increasing $\beta$ favors axial instabilities that eventually develop into the square skyrmion crystal observed in the Monte Carlo simulations.

By contrast, the ESP interaction contributes only through fourth-order momentum terms. Consequently, it does not alter the leading instability of the polarized state, but rather acts as a higher-order correction whose influence becomes significant only once finite-wavevector modulated states are already established.

\subsubsection{Implications for the square skyrmion crystal}

The previous analysis provides a natural explanation for the behavior of the BP model. Since the $\beta^2$ interaction modifies the quadratic dispersion of the low-energy fluctuations, increasing $\beta$ directly favors axial instabilities of the polarized state, which eventually evolve into the square skyrmion crystal observed in the Monte Carlo phase diagram.

The situation is qualitatively different for the ESP model. As shown above, the leading contribution generated by the $\beta^2$ interaction appears only at fourth order in the momentum expansion and therefore does not modify the leading instability of the polarized phase. This is consistent with our numerical results, where the square skyrmion crystal does not emerge directly from the ferromagnetic state but rather through a reconstruction of an already established triangular skyrmion crystal.

This observation also indicates that the continuum expansion around the polarized state is no longer sufficient to understand the ESP mechanism. Once a multi-$\mathbf q$ texture is formed, the phonon-induced interaction becomes intrinsically nonlinear, coupling different finite-wavevector modes of the magnetic texture. Consequently, the relative stability of triangular and square skyrmion crystals cannot be inferred from the quadratic fluctuation spectrum alone.

The origin of this difference can be traced back to the order in spatial gradients of the effective interactions. In the BP model, the $\beta^2$ contribution modifies the quadratic fluctuation spectrum already at order $q^2$, and therefore changes the instability of the polarized state itself. In contrast, the ESP contribution enters at order $q^4$. Its effect is weaker in the long-wavelength expansion around the polarized state, but it can still modify the relative energy of finite-amplitude multi-$\mathbf q$ textures.

To gain qualitative insight into this competition, we consider simple variational representations of triangular and square skyrmion crystals obtained from the superposition of three and two helices, respectively. For simplicity, we neglect the local hard-spin normalization and introduce an auxiliary magnetization field
\begin{eqnarray}
\mathbf n(\mathbf r)&=&
\left[
n_0+
A_z\sum_{\nu=1}^{N_q}
\cos(\mathbf Q_\nu\cdot\mathbf r+\phi_\nu)
\right]\hat{\mathbf z}
\nonumber\\
&&+
A_\perp\sum_{\nu=1}^{N_q}
\hat{\mathbf e}_{\nu}
\sin(\mathbf Q_\nu\cdot\mathbf r+\phi_\nu),
\label{Eq:ansatz_general}
\end{eqnarray}
where
\begin{equation}
\hat{\mathbf e}_{\nu}
=
\hat{\mathbf z}\times\hat{\mathbf Q}_{\nu},
\end{equation}
and $n_0$, $A_z$, and $A_\perp$ are the variational amplitudes
defined in the main text. This approximation is used only to identify the energetic tendency of the phonon-induced interaction and not as a complete variational solution of the hard-spin model.

The square skyrmion crystal is represented by a two-$\mathbf q$ state ($N_q=2$), with
\begin{equation}
\mathbf Q_1=(Q,0),
\qquad
\mathbf Q_2=(0,Q).
\label{Eq:Qsquare}
\end{equation}
For this simple comparison, the corresponding phases can be set to zero.

The triangular skyrmion crystal is represented by a three-$\mathbf q$ state ($N_q=3$), with
\begin{eqnarray}
\mathbf Q_1&=&Q(1,0),\nonumber\\
\mathbf Q_2&=&Q\left(-\frac{1}{2},\frac{\sqrt{3}}{2}\right),\nonumber\\
\mathbf Q_3&=&Q\left(-\frac{1}{2},-\frac{\sqrt{3}}{2}\right),
\label{Eq:Qtriangular}
\end{eqnarray}
which satisfy
\begin{equation}
\mathbf Q_1+\mathbf Q_2+\mathbf Q_3=0.
\end{equation}
In this case, the phonon-induced energy depends on the phase combination
\begin{equation}
\Phi=\phi_1+\phi_2+\phi_3.
\end{equation}

Within this simplified analysis, we evaluate only the additional $\beta^2$ contribution. This isolates the energetic bias introduced by the spin-phonon coupling, without attempting to determine the full energy balance or the exact phase boundaries, which also depend on exchange, bilinear Dzyaloshinskii-Moriya interaction, magnetic field, and the local hard-spin constraint.

Substituting the two ans\"atze into Eqs.~(\ref{Eq:FBP}) and (\ref{Eq:FESP}) and averaging over one magnetic unit cell allows us to compare how the phonon-induced interaction acts on the square and triangular multi-$\mathbf q$ textures.

For the square state, choosing the phases equal to zero, Eq.~(\ref{Eq:ansatz_general}) gives
\begin{eqnarray}
n_x(\mathbf r)&=&-A_\perp\sin(Qy),\nonumber\\
n_y(\mathbf r)&=& A_\perp\sin(Qx),\nonumber\\
n_z(\mathbf r)&=&n_0+A_z\left[\cos(Qx)+\cos(Qy)\right].
\label{Eq:square_ansatz}
\end{eqnarray}

For the BP interaction, the two directional chiral combinations entering Eq.~(\ref{Eq:FBP}) become
\begin{eqnarray}
n_y\partial_xn_z-n_z\partial_xn_y&=&-A_\perp Q\left[
A_z+n_0\cos(Qx)
\right.\nonumber\\
&&\left. +\,
A_z\cos(Qx)\cos(Qy)
\right],
\nonumber\\
n_z\partial_yn_x-n_x\partial_yn_z
&=&
-A_\perp Q
\left[
A_z+n_0\cos(Qy)
\right.\nonumber\\
&&\left. +\,
A_z\cos(Qx)\cos(Qy)
\right].
\end{eqnarray}

Averaging their squared sum over one magnetic unit cell gives
\begin{equation}
\overline{\mathcal F}_{\beta,\square}^{\rm BP}
=
-K_{\rm BP}A_\perp^{\,2}Q^2
\left(
n_0^2+\frac{5}{2}A_z^2
\right).
\label{Eq:EsquareBP}
\end{equation}

For the ESP interaction, the corresponding second-derivative combinations are
\begin{eqnarray}
n_y\partial_x^2n_z-n_z\partial_x^2n_y
&=&
A_\perp Q^2
\left[
n_0+A_z\cos(Qy)
\right]\nonumber\\
&&\quad\times\sin(Qx),
\nonumber\\
n_z\partial_y^2n_x-n_x\partial_y^2n_z
&=&
A_\perp Q^2
\left[
n_0+A_z\cos(Qx)
\right]\nonumber\\
&&\quad\times\sin(Qy).
\end{eqnarray}

Their spatial average yields
\begin{equation}
\overline{\mathcal F}_{\beta,\square}^{\rm ESP}
=
-K_{\rm ESP}A_\perp^{\,2}Q^4
\left(
n_0^2+\frac{1}{2}A_z^2
\right).
\label{Eq:EsquareESP}
\end{equation}

These expressions again show the different momentum dependence of the two models. The BP energy gain scales as $Q^2$, while the ESP contribution scales as $Q^4$. In both cases, however, the two axial helices contribute coherently to the phonon-induced energy of the square texture.

For the triangular state, using the ordering vectors in
Eq.~(\ref{Eq:Qtriangular}), the spatially averaged BP contribution is

\begin{equation}
\overline{\mathcal F}_{\beta,\triangle}^{\rm BP}
=
-K_{\rm BP}A_\perp^{\,2}Q^2
\left[
\frac{9}{8}n_0^2
+
\frac{117}{16}A_z^2
+
\frac{9}{8}n_0A_z\cos\Phi
\right],
\label{Eq:EtriangleBP}
\end{equation}

whereas the corresponding ESP contribution is

\begin{equation}
\overline{\mathcal F}_{\beta,\triangle}^{\rm ESP}
=
-K_{\rm ESP}A_\perp^{\,2}Q^4
\left[
\frac{15}{16}n_0^2
+
\frac{9}{16}A_z^2
+
\frac{9}{16}n_0A_z\cos\Phi
\right].
\label{Eq:EtriangleESP}
\end{equation}

The phase dependence in Eqs.~(\ref{Eq:EtriangleBP}) and
(\ref{Eq:EtriangleESP}) originates from the condition
$\mathbf Q_1+\mathbf Q_2+\mathbf Q_3=0$, which allows the three helical
components to interfere coherently. For the usual phase relation of the
triangular skyrmion crystal, $\Phi=\pi$, the ESP energy gain of the
triangular state becomes

\begin{equation}
\overline{\mathcal F}_{\beta,\triangle}^{\rm ESP}
=
-K_{\rm ESP}A_\perp^{\,2}Q^4
\left[
\frac{15}{16}n_0^2
+
\frac{9}{16}A_z^2
-
\frac{9}{16}n_0A_z
\right].
\label{Eq:EtriangleESPpi}
\end{equation}

Comparing this expression with Eq.~(\ref{Eq:EsquareESP}), the difference
between the corresponding ESP energy gains can be written as

\begin{equation}
G_{\square}^{\rm ESP}
-
G_{\triangle}^{\rm ESP}
=
\frac{1}{16}
\left(
n_0^2-A_z^2+9n_0A_z
\right),
\label{Eq:ESP_energy_difference}
\end{equation}

where
$\overline{\mathcal F}_{\beta}^{\rm ESP}
=-K_{\rm ESP}A_\perp^{\,2}Q^4G^{\rm ESP}$.
For a finite positive uniform magnetization and comparable values of
$n_0$ and $A_z$, the right-hand side of
Eq.~(\ref{Eq:ESP_energy_difference}) is positive. Within this simplified
ansatz, the phonon-induced $\beta^2$ interaction can therefore provide a
larger energy gain for the square two-$\mathbf q$ texture than for the
triangular three-$\mathbf q$ state.

This simple comparison provides a possible interpretation of the
triangular-to-square reconstruction observed in the ESP phase diagram.
The higher-order ESP interaction does not generate the initial
finite-$\mathbf q$ instability of the polarized state. Instead, once a
multi-$\mathbf q$ skyrmion texture is present, it modifies the nonlinear
coupling between its helical components and can bias the energetic
competition in favor of the square two-$\mathbf q$ arrangement.

We stress that this argument is not intended as a complete variational
determination of the phase boundary. The auxiliary field in
Eq.~(\ref{Eq:ansatz_general}) is not locally normalized, and the complete
energy balance also involves exchange, bilinear Dzyaloshinskii-Moriya
interaction, magnetic field, higher harmonics, and the hard-spin
constraint. Nevertheless, the continuum and variational analyses
illustrate why the two phonon models can stabilize square skyrmion
crystals through different mechanisms: the BP interaction modifies the
leading instability of the polarized phase, whereas the ESP interaction
mainly changes the relative energy of already established
multi-$\mathbf q$ textures.

%\bibliography{biblio}

%merlin.mbs apsrev4-1.bst 2010-07-25 4.21a (PWD, AO, DPC) hacked
%Control: key (0)
%Control: author (0) dotless jnrlst
%Control: editor formatted (1) identically to author
%Control: production of article title (0) allowed
%Control: page (1) range
%Control: year (0) verbatim
%Control: production of eprint (0) enabled
%

\end{document}